\documentclass[twocolumn]{aastex62}
\usepackage{textcomp}
\usepackage{amsmath} 
\usepackage{wasysym}
\usepackage{graphicx}
\usepackage{amssymb}
\usepackage{epstopdf}
\usepackage{mathrsfs}
\usepackage{anyfontsize}
\usepackage{natbib}
\usepackage{placeins}
\usepackage{color}
\usepackage{lipsum}
\usepackage{diagbox}
\usepackage{gensymb}
\usepackage{booktabs}
\usepackage{appendix}
\usepackage{graphicx}
\usepackage[left]{lineno} % 里面的选项代表双栏 switch

\usepackage{CJKutf8}
\usepackage{bm}

\shorttitle{structure growth and SF in LSB and HSB}
\shortauthors{Shen et al.}
\begin{document}
\begin{CJK}{UTF8}{gbsn}	
    
    \title{SDSS-IV MaNGA: Distinct Structural Growth and Star Formation in Low and High Surface Brightness Disks}
    \correspondingauthor{Jun Yin, Lei Hao, Chong Ge}
    \email{jyin@shao.ac.cn, haol@shao.ac.cn, chongge@xmu.edu.cn}
     
    \author[0009-0003-2072-1200]{Mengting Shen}
    \affiliation{Department of Astronomy, Xiamen University, Xiamen, Fujian 361005, China}
    
    \author[0000-0002-4499-1956]{Jun Yin}
    \affiliation{Key Laboratory for Research in Galaxies and Cosmology, Shanghai Astronomical Observatory, Chinese Academy of Sciences, 80 Nandan Road, Shanghai 200030, China}
    
    \author[0000-0002-4176-9145]{Hassen M. Yesuf}
    \affiliation{Key Laboratory for Research in Galaxies and Cosmology, Shanghai Astronomical Observatory, Chinese Academy of Sciences, 80 Nandan Road, Shanghai 200030, China}

    \author[0000-0003-2478-9723]{Lei Hao}
    \affiliation{Key Laboratory for Research in Galaxies and Cosmology, Shanghai Astronomical Observatory, Chinese Academy of Sciences, 80 Nandan Road, Shanghai 200030, China}
    
    \author[0000-0002-8817-4587]{Jiafeng Lu}
    \affiliation{Institute for Astronomy, School of Physics, Zhejiang University, Hangzhou 310027, China}
    
    \author[0000-0003-1138-8146]{Lin Lin}
    \affiliation{Key Laboratory for Research in Galaxies and Cosmology, Shanghai Astronomical Observatory, Chinese Academy of Sciences, 80 Nandan Road, Shanghai 200030, China}
    
    \author[0000-0003-0628-5118]{Chong Ge}
    \affiliation{Department of Astronomy, Xiamen University, Xiamen, Fujian 361005, China}
    
    \author[0000-0003-4874-0369]{Junfeng Wang}
    \affiliation{Department of Astronomy, Xiamen University, Xiamen, Fujian 361005, China}

    \author[0000-0002-3073-5871]{Shiyin Shen}
    \affiliation{Key Laboratory for Research in Galaxies and Cosmology, Shanghai Astronomical Observatory, Chinese Academy of Sciences, 80 Nandan Road, Shanghai 200030, China}
    
    \author[0000-0002-2204-6558]{Yu Rong}
    \affiliation{Department of Astronomy, University of Science and Technology of China, Hefei, Anhui 230026, People’s Republic of China}
    \affiliation{School of Astronomy and Space Sciences, University of Science and Technology of China, Hefei 230026, Anhui, People’s Republic of China}
    
    \begin{abstract}
    We analyze a clean sample of 1,118 late-type, face-on galaxies without AGN contamination from the MaNGA survey. Their photometric structures are quantified via two-component (bulge+disk) decompositions on deep $g$-band images from the DESI Legacy Survey. Using a disk central surface brightness of $\mu_{\rm 0,d,cor}$(g) = 22 ± 0.3 mag arcsec$^{-2}$ (corrected for inclination and cosmic dimming) as the classification threshold, we identify 159 low surface brightness (LSB) galaxies, 388 LSB candidates, and 571 high surface brightness (HSB) galaxies.
    LSB galaxies are predominantly low-mass ($M_\ast < 3 \times 10^{10}$ M$_\odot$), exhibiting 29\% larger effective radii, 15\% lower star formation rates (SFRs), and 12\% reduced gas-phase metallicities than HSB counterparts at comparable masses. These differences cause systematic offsets from standard scaling relations. Despite comparable gas content, LSB galaxies host older stellar populations, longer gas depletion times, and less efficient star formation.
    Spatially resolved analyses further reveal that LSB galaxies display centrally suppressed $\Sigma_{\rm SFR}$, flatter SFR gradients, and rising specific SFR profiles toward their outskirts. Together with steeper negative metallicity gradients, these trends suggest ongoing gas accretion fueling outer-disk star formation. Consistently, the outer regions of LSB galaxies exhibit stronger H$\delta_A$ absorption and lower D$_n$4000 indices, indicating fading A-star populations. Moreover, LSB galaxies show lower $\Sigma_{\ast}$ across all $R/R_e$ and more centrally depleted stellar mass profiles on an absolute radial scale, compared with HSB and large-size star-forming galaxies. Collectively, LSB galaxies represent a distinct population with slow evolution, inefficient star formation, and continued susceptibility to late-time gas accretion and peripheral star formation.
    \end{abstract}
    
    \keywords{galaxies:low surface brightness, galaxies:fundamental parameters, galaxies:star formation, galaxies:disc}
    
    % \linenumbers
    
    \section{INTRODUCTION}
    \label{sec:INTRODUCTION}
    
    LSB galaxies are a class of galaxies characterized by a central surface brightness that is at least one magnitude fainter than the night-sky background at a site without light pollution \citep{Impey_1997, Vollmer_2013}. Observationally, classical disk galaxies follow the Freeman Law, which refers to disks and provides a characteristic face-on central surface brightness of $\mu_0(B)$= 21.65$\pm$0.3 mag arcsec$^{-2}$ \citep{Freeman_1970}. Galaxies with disk central surface brightness fainter than $\mu_{\rm 0,d}(B)=22.5$ mag arcsec$^{-2}$ \citep{McGaugh_s_1996, Rosenbaum_2009, Du_2015}—approximately one magnitude fainter than the Freeman value—are therefore classified as LSB galaxies, while brighter counterparts are defined as HSB galaxies. Although they contribute only a few percent of the local luminosity and stellar mass density (e.g., \citeauthor{Bernstein_1995} \citeyear{Bernstein_1995}; \citeauthor{Hayward_2005} \citeyear{Hayward_2005}; \citeauthor{Martin_2019} \citeyear{Martin_2019}), making them difficult to detect, LSB galaxies may comprise up to 20\% of the total dynamical mass in the universe, including dark matter and gas \citep{Minchin_2004}. Furthermore, LSB galaxies could contribute between 30\% and 60\% to the number density of local galaxies \citep{McGaugh_1996, Trachternach_2006, Haberzettl_2007}.
    
    There are several indications that LSB galaxies exhibit a range of colors, with the most massive ones appearing blue \citep{Impey_1997, Zackrisson_2005, Vorobyov_2009}, while some very extended dwarfs, such as ultra-diffuse galaxies (UDGs), which are beyond the scope of this study, may appear red \citep{Leisman_2017, Eigenthaler_2018, Rong_2020}. In comparison to HSB galaxies, LSB galaxies appear to have low metallicity \citep{Galaz_2006, Liang_2010}, with oxygen abundances as low as approximately 1/3 to 1/5 of solar abundance \citep{Roennback_1995, Burkholder_2001}. Furthermore, they possess higher $M_\ast$/L ratios \citep{Sprayberry_1995}, reduced dust mass \citep{Matthews_2001}, greater amounts of dark matter \citep{de_Blok_1997}, and higher spin parameters \citep{Leisman_2017, Hua_2025}. Despite containing substantial gas reservoirs \citep{Mo_1994, Boissier_2008, Schombert_2011}, their gas surface densities are about 1/3 lower than HSB galaxies \citep{de_Blok_1996, Gerritsen_1999}, falling below the critical threshold for efficient star formation \citep{Kennicutt_1989, van_der_Hulst_1993, Lei_2019} , and resulting in significantly lower star formation rates \citep{O'Neil_2007, Schombert_2011, Rong_2020}.
    
    The mechanisms behind this suppressed star formation remain an active area of research. Studies suggest multiple contributing factors: the low conversion efficiency of H$\,\textsc{i}$ to $\rm H_2$ gas \citep{Cao_2017}, with molecular gas accounting for only about 3\% of the cold gas in late-type LSB galaxies \citep{McGaugh_2017}. Interestingly, while LSB galaxies exhibit H$\,\textsc{i}$ surface densities comparable to star-forming galaxies, their SFRs fall well below predictions from the Schmidt law \citep{Lei_2018, Lei_2019}. This discrepancy suggests that environmental isolation alone cannot explain their low star formation activity, as LSB galaxies are typically found in low-density regions yet remain gas-rich \citep{Galaz_2011, Du_2015, Honey_2018, Perez_Montano_2019}.
    
    Recent simulations have significantly advanced our understanding of LSB galaxy formation and evolution. The Horizon-AGN simulation reveals that LSB and HSB galaxies likely share a common origin, with LSB systems experiencing rapid early star formation followed by reduced activity after $z \sim 1$, to evolve slowly \citep{Martin_2019, Stoppacher_2025}. Tidal interactions during this period expanded their stellar distributions and heated cold gas, creating the diffuse systems we observe today. Further studies using SDSS data and TNG100 simulations demonstrate that LSB galaxies possess higher specific stellar angular momentum and halo spin than HSB galaxies \citep{Perez_Montano_2019, Perez_Montano_2022}. This leads to more extended structures that inhibit gas inflow to central regions, effectively suppressing star formation. \citet{Stoppacher_2025} confirm these findings through hydrodynamical and EAGLE simulations, identifying angular momentum as the key driver of differences between LSB and HSB evolutionary paths.
    
    The formation mechanisms of LSB galaxies appear to be mass-dependent. For low-mass systems, stellar and AGN feedback mechanisms may play a significant role \citep{Rong_2017, Christensen_2018, Di_Cintio_2019}, while more massive LSB galaxies may form through mergers \citep{Saburova_2018, Di_Cintio_2019, Zhu_2023}, gas accretion \citep{Saburova_2021}, or bar-driven processes \citep{Noguchi_2001}. Additionally, classical formation models indicate that LSB galaxies reside in halos with relatively high spin parameters \citep{Mo_1998, Rong_2017, Kim_2013, Hua_2025}. These diverse formation pathways help explain the varied properties observed in LSB galaxies, including elevated gas fractions, high spin parameters, and suppressed star formation rates.
    
    Despite numerous studies on star formation in LSB galaxies, the reasons for their lower star formation rates remain poorly understood, even with large samples \citep{zhong_2008,Du_2015}. Earlier research often relied on central surface brightness of single-component fitting of galaxies to identify LSB galaxies, with subsequent studies focusing on late-type disk-dominated LSB galaxies \citep{McGaugh_1994, de_Blok_1995, zhong_2008, Du_2015} or individual giant galaxies like Malin-type LSB galaxies \citep{Sprayberry_1995, Pickering_1997}. Recently, the advent of multi-component decomposition catalogs for large galaxy samples, derived from bulge-disk decomposition, enables selection of not only disk-dominated LSB galaxies but also those hosting bulge components, enhancing our understanding of their properties. Furthermore, the rise of Integrated Field Spectroscopy (IFS) surveys offers unprecedented spatial resolution, enabling detailed investigations of star formation gradients both in the central regions and outer disk regions of LSB galaxies, areas previously underexplored in the literature.
    
    In this study, we used the Mapping Nearby Galaxies at Apache Point Observatory (MaNGA) integrated spectroscopic survey, combining it with careful galaxy structure decomposition using the GALFIT software \citep{Peng_2002, Peng_2010}, based on deep Dark Energy Spectroscopic Instrument (DESI) Legacy $g$-band images. 
    Classically, LSB galaxies are defined by the central surface brightness of the disk component in the B-band, where $\mu_{\rm 0,d}(B) \geq$ 22.5 mag arcsec$^{-2}$ \citep{McGaugh_s_1996, Rosenbaum_2009, Du_2015}. Although the B-band value can be converted from $g$ and $r$ band photometry using established color conversion formulas \citep{zhong_2008, Pahwa_2018}, bulge-disk decomposition results are sensitive to the bandpass——fitting in different bands may yield varying component values that impact classification robustness. To address this and ensure consistency, we adopt a $g$-band equivalent threshold directly: $\mu_{\rm 0,d,cor}(g) =$ 22 ± 0.3  mag arcsec$^{-2}$ \citep{Tanoglidis_2021}, which is calibrated to match the B-band standard and corrected for inclination and cosmic dimming effects. Based on this corrected threshold, we categorize galaxies into three samples: LSB galaxies ($\mu_{\rm 0,d,cor}$(g) $\geq$ 22.3 mag arcsec$^{-2}$), LSB candidates (21.7 $< \mu_{\rm 0,d,cor}$(g) $<$ 22.3 mag arcsec$^{-2}$), and HSB galaxies ($\mu_{\rm 0,d,cor}$(g) $\leq$ 21.7 mag arcsec$^{-2}$). We then explored the global and resolved star formation and metallicity characteristics of LSB galaxies using MaNGA data, further examining their gas properties through correlations with the HI-MaNGA catalog \citep{Masters_2019}. This enables us to characterize the star formation properties of the LSB galaxies.
    %For our analysis, this B-band threshold serves as the fundamental selection criterion, with equivalent values derived from g- and r-band photometry where direct B-band measurements are unavailable \citep{zhong_2008, Pahwa_2018}. Following bulge-disk decomposition, we categorized galaxies by the $g$-band disk central surface brightness, $\mu_{\rm 0,d,cor}$(g) = 22 ± 0.3 mag arcsec$^{-2}$ \citep{Tanoglidis_2021}, corrected for inclination and cosmic dimming. This correction was calibrated to match the B-band standard, defining three samples: 
    
    We describe our data in Section \ref{sec:DATA}. The sample is introduced in Section \ref{sec:THE SAMPLE}, where we also outline the detailed process of two-component decomposition and the method of selecting LSB samples. The results are discussed in Section \ref{sec:RESULTS}, which mainly includes the global properties of our selected LSB sample as well as radial profile and gradient. Our understanding of these results is provided in Section \ref{sec:DISCUSSION}. Finally, a summary of our work is presented in Section \ref{sec:SUMMARY AND CONCLUSION}. Throughout this paper, we adopt a set of cosmological parameters as follows: $H_0$=70 km s$^{-1}$ Mpc$^{-1}$, $\Omega_{m}$=0.30, $\Omega_{\Lambda}$=0.70.
    
    \section{DATA}
    \label{sec:DATA}
    
    \subsection{The MaNGA Data}
    
    MaNGA is a sky survey that began in July 2014 and is one of the three core projects of the SDSS-IV \citep{Bundy_2015, Law_2016}. This survey deploys 17 pluggable integrated Field of View units (IFUs), each comprising hexagonal fiber bundles that contain between 19 to 127 fibers. These units offer varying field-of-view sizes from $12\overset{\arcsec}{.}5$ to $32\overset{\arcsec}{.}5$, enabling the collection of 3D spectra with a spatial resolution of 2$\arcmin$ \citep{Drory_2015, Law_2016}. The MaNGA spectrum covers a wavelength range from 3,600 \AA\;to 10,000 \AA\ with a spectral resolution of R$\sim$2000. This capability allows for the precise measurement of all strong emission lines present within the wavelengths ranging from [O$\,\textsc{ii}$]$\lambda$3727 to [S$\,\textsc{ii}$]$\lambda$6731 \citep{Bundy_2015}. Consequently, the survey has successfully acquired integrated field-of-view spectra for approximately 10,000 nearby galaxies. These galaxies have stellar masses between 10$^9$-10$^{11}$M$_\odot$ and redshifts ($z$) within the range of 0.01 $< z <$ 0.15 \citep{Blanton_2017}. For details on the selection criteria of the MaNGA sample, please refer to \citet{Wake_2017}.%$0\overset{\arcsec}{.}5$ 

    In this study, we use the latest 11th internal data release from MaNGA, known as the MaNGA Product Release 11 (MPL-11), which includes 10,782 unique galaxies. This dataset is also included in the upcoming SDSS Data Release 17 (DR17; SDSS collaboration, in prep.). The analysis employs both the Data Analysis Pipeline (DAP; \citeauthor{Westfall_2019} \citeyear{Westfall_2019}; \citeauthor{Belfiore_2019} \citeyear{Belfiore_2019}) and the Data Reduction Pipeline (DRP; \citeauthor{Law_2016} \citeyear{Law_2016}). These pipelines facilitate the spectral fitting of the MaNGA data, generating two-dimensional spatial distribution information for various physical characteristics, including star formation rate, gas phase metallicity, emission line intensity, and spectral index. This work focuses on examining the radial profiles of the aforementioned parameters based on the MaNGA dataset, providing insights into the spatial and physical properties of the observed galaxies.

    \subsection{Imaging Data from the DESI}
    \label{sec:Imaging data from the DESI}
    
    The DESI Legacy Imaging Survey \citep{Dey_2019} combines three wide-field optical imaging surveys: the Dark Energy Camera Legacy Survey (DECaLS; \citeauthor{Flaugher_2015} \citeyear{Flaugher_2015}), the Mayall z-band Legacy Survey (MzLS; \citeauthor{Dey_2016} \citeyear{Dey_2016}), and the Beijing-Arizona Sky Survey (BASS; \citeauthor{Williams_2004} \citeyear{Williams_2004}). Collectively, these surveys cover an impressive total area of approximately 14,000 deg$^2$. In the northern sky, BASS provides imaging in the $g$ and $r$ bands, while MzLS focuses on the $z$ band, achieving 5$\sigma$ detection limits of $g$ = 23.48, $r$ = 22.87, and $z$ = 22.29 AB mag. Conversely, the southern sky is mapped by DECaLS across all three bands ($g$, $r$, and $z$), with 5$\sigma$ detection limits of $g$ = 23.72, $r$ = 23.27, $z$ = 22.22 AB mag. Notably, these detection limits were determined for a fiducial galaxy size of $0\overset{\arcsec}{.}45$. For this analysis, we use publicly available data from DESI DR9 and DR10, both of which feature enhanced restoration techniques and procedures compared to the earlier DR8 data. The initial release includes images and catalogs from all three Legacy Surveys, while DR10 further enriches the dataset by incorporating additional DECam data from NOIRLab in previously unobserved regions. Therefore, our focus will be on the latest $g$-band image data sourced from both DR9 and DR10.
    
    \section{THE SAMPLE}
    \label{sec:THE SAMPLE}
    
    \subsection{Parent Sample of Disk and Face-on Galaxies}
    
    To better understand the formation and evolution of LSB galaxies, we used the parent samples of late-type galaxies selected from the MaNGA MPL-11 release data. This dataset encompasses 9,853 galaxies with high-quality two-dimensional spectral observations. We supplemented the MaNGA data with additional catalogs. Specifically, we obtained the bulge-to-disk ratio (B/T) from the MaNGA PyMorph photometric Value Added Catalogue (MPP-VAC) catalog \citep{Sanchez_2022} and the morphological type (T-type) parameters from the MDLM-VAC catalog \citep{Sanchez_2022}. The global stellar mass ($M_\ast$) and dust attenuation in rest-frame V ($A_V$) were sourced from the GSWLC-A2 catalog, while the SFR was calculated through UV/optical SED fitting \citep{Salim_2016, Salim_2018}. Other intrinsic properties, including redshift ($z$), half-light radius ($R_{50}$), axis ratio ($q$ = $b/a$), and photometric major axis position angle ($\phi$), were obtained from the NASA-Sloan Atlas (NSA, \citeauthor{Blanton_2011} \citeyear{Blanton_2011}). Furthermore, gas-phase metallicity (12+log(O/H)) and luminosity-weighted age indicators such as $D_{n}4000$ were derived from the integrated data within one effective radius (half-light radius, $R_e$) of the MaNGA survey \citep{Westfall_2019}.

    Combining all this information, we defined a set of criteria to select a parent sample of nearly face-on, disk-dominated late-type galaxies:
    
    (i) $b/a>$0.7. To select face-on galaxies.
    
    (ii) log([O$\,\textsc{iii}$]/H$\beta$) $<$ 0.61/(log([N$\,\textsc{ii}$]/H$\alpha$)-0.47)+1.19. Based on the partition method proposed by \citet{Kauffmann_2003}, we select galaxies whose central pixels are not classified as active galactic nuclei (AGN).
    
    (iii) T-Type$>$0 and S\'ersic $n<$4. Late-type galaxies are identified according to the MDLM-VAC catalog.
    
    (iv) B/T$<$0.5 and $n_{\rm disk} \neq$ -999. Select galaxies with non-dominant bulges and successful disk decomposition.
    
    Ultimately, using these criteria, we selected a total of 1,223 galaxies as our parent sample.
    
    \subsection{Multi-Model Bulge-Disk Decomposition}
    \label{sec:Multi-model bulge-disk decomposition}
     
    A key parameter for classifying a galaxy as low (or high) surface brightness is the corrected central surface brightness of its disk component, denoted as $\mu_{\rm 0, d, cor}$ which accounts for inclination and cosmic dimming effects (see Section \ref{sec:Selection of LSB galaxies} for details). In other words, the classification heavily depends on the results of component decomposition. Although the MPP-VAC catalog \citep{Sanchez_2022} provides two-component decomposition (see details in \citeauthor{Fischer_2019} \citeyear{Fischer_2019}; \citeauthor{Sanchez_2022} \citeyear{Sanchez_2022}), for LSB galaxies the decompositions may not be reliable. Therefore, in this work, we use the $g$-band fitting results from the MPP-VAC catalog as initial parameters to refit magnitudes, effective radii, axial ratios, and position angles of the bulge and disk components using the deeper DESI Legacy Survey images.
    
    We used GALFIT software \citep{Peng_2002, Peng_2010} to perform 2D bulge-disk decomposition for each galaxy image. The software fits two-component (bulge + disk) models to the light distributions, optimized to achieve the best fit to the galaxy morphology. In this study, we apply two models to the light distribution in the $g$-band: S\'ersic bulge + Exponential disk (\textsc{SerExp}) and S\'ersic bulge + S\'ersic disk (\textsc{SerSer}). The specific formula of the S\'ersic function can be referred to Eq. \ref{formula2}. It is worth noting that about 180 galaxies in the parent sample exhibit a barred structure. Ignoring this bar component could lead to an overestimation of the central surface brightness for the decomposed disk component. However, when focusing solely on the bulge and disk components, the currently selected LSB galaxies remain their LSB classification, although this approach may slightly undercount the total number of true LSB galaxies. Therefore, this study focuses on the two-dimensional bulge-disk decomposition of galaxy images.

\begin{figure*}[htbp]%
    \centering
    \includegraphics[width=0.95\textwidth]{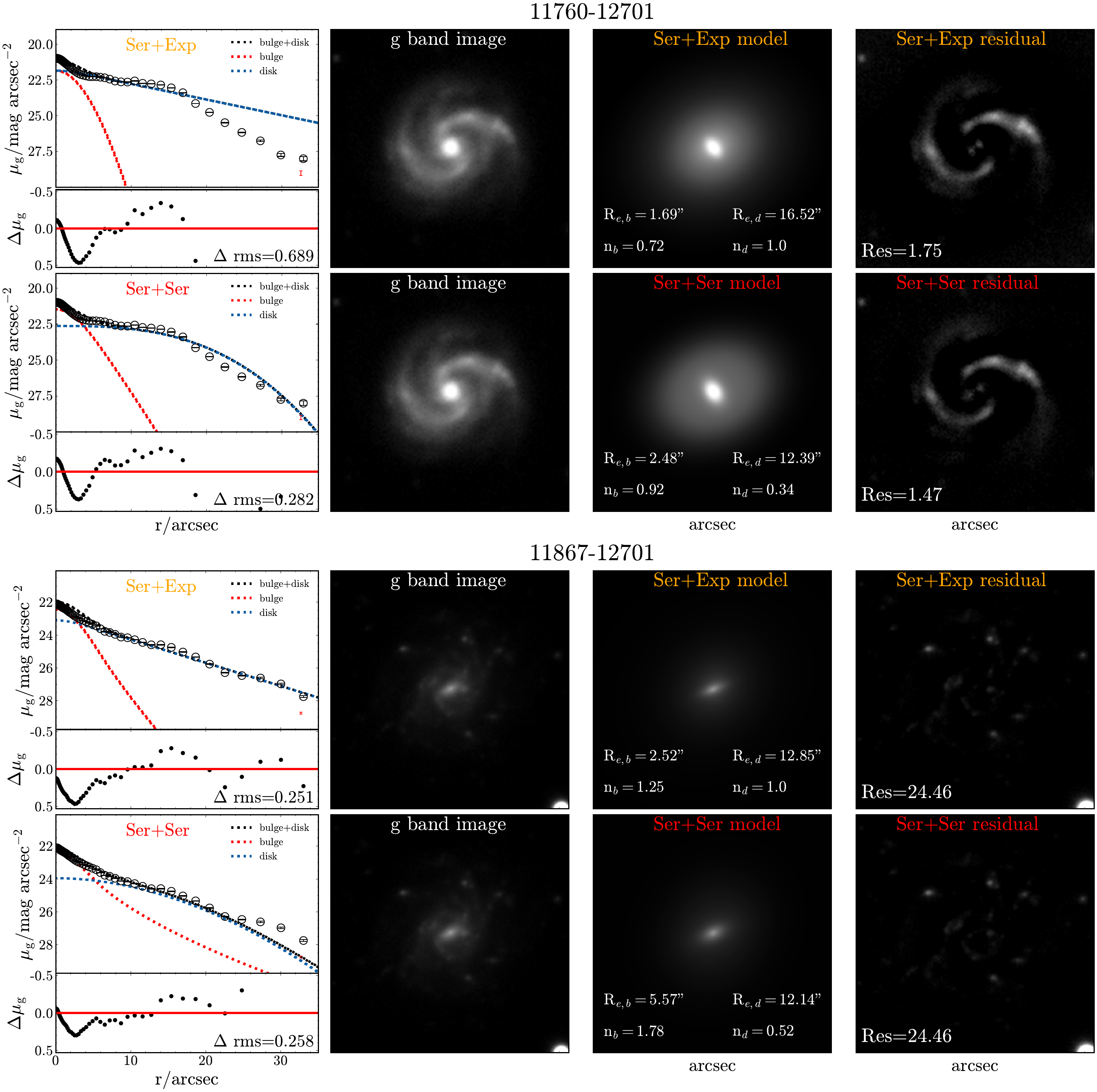} 
    \caption{The two-component decomposition results for galaxies 11760-12701 (a bulge-dominated galaxy with spiral arms, upper eight panels) and 11867-12701 (a disk-dominated galaxy with weak spiral arms, lower eight panels). The first and third rows display decomposition results for a S\'ersic bulge combined with an exponential disk, while the second and fourth rows show results for a S\'ersic bulge combined with a S\'ersic disk. The first column presents fitting results for the surface brightness radial profile and corresponding residuals, where black open circles represent the median $g$-band surface brightness obtained from elliptical isophotal fitting of the galaxies; red and blue dotted lines are the fitting results for the bulge and disk components, respectively, and the black dotted line indicates the fitting result for the total surface brightness of the bulge and disk. All model lines are convolved with the PSF, and the maximum error of the observed data points (red error bar) is labeled in the lower right corner of each panel. The residuals $\Delta\mu_g$ (black dots) are calculated as the difference between the measured surface brightness (black open circles) and the total surface brightness fitting curve (black dotted line). To guide the eye, the horizontal red line indicates a 0 mag arcsec$^{-2}$ difference between fit and data, while the $\Delta$rms of the deviation between fit and data is given in the lower right corner. The second column presents $g$-band images from the DESI Survey. The third column shows model images with annotations of $R_e$ and S\'ersic index $n$ for each component (bulge and disk). The fourth column displays residual images, with the sum of residuals for each image labeled in the lower left corner, and its unit is nanomaggy.}
    \label{fig1}
    \end{figure*}

    The images to be fitted and the PSF images required by the GALFIT software are taken from the DESI survey. Each galaxy image is cropped to a size of 240×240 pixel$^2$ (equivalent to 62.88 × 62.88 arcsec$^2$). This size ensures that the entire galaxy is fully enclosed while preserving a sufficient surrounding sky background area to facilitate accurate sky background estimation. Source Extract Python (SEP) is employed to create a mask image for each galaxy, effectively removing surrounding sources in the vicinity of the target galaxy within the image. Concurrently, it is necessary to input the initial parameters of the bulge and disk components, including the galactic center, component magnitude, half-light radii (or scale length), $b/a$ ratio, and position angle. These initial parameter values are obtained from the fitted values in the $g$-band image provided by the MPP-VAC catalog \citep{Sanchez_2022}. After setting the initial parameters, we used GALFIT to fit the $g$-band images of all 1,223 galaxies in both \textsc{SerExp} and \textsc{SerSer} modes. The output of the GALFIT fit comprises three images, namely the input image, the final model, and the residual image, which is created by subtracting the final model from the original image. In Fig. \ref{fig1}, we present the decomposition results of the two-component \textsc{SerExp} and \textsc{SerSer} for galaxies 11760-12701 (bulge-dominated galaxy with spiral arms) and 11867-12701 (disk-dominated galaxy with weak spiral arms) as examples.
    
    While each fitting result provides a $\chi^2$ value to characterize the goodness of fit, these values are not suitable for comparing different models. A model with more fitting parameters will naturally fit the given data better. This additional flexibility must be considered when assessing whether more complex models have statistical support. The Bayesian Information Criteria (BIC) introduces penalties for models with increased complexity. BIC is expressed as follows:

    \begin{equation}
	    BIC=\chi^2+k*ln(n)
    \label{formula1}
    \end{equation}
    Where $\chi^2$ represents the chi-square statistic, $k$ denotes the number of free parameters in the various fitting functions (with $k=13$ for \textsc{SerExp}, $k=14$ for \textsc{SerSer}), and $n$ is the number of data points in the fitted image. A smaller BIC value indicates a better-fitting model. By comparing the BIC values of different models, we select the most suitable two-component fitting model for each galaxy. As shown in Fig. \ref{fig1}, galaxy 11760-12701 is a bulge-dominated spiral galaxy. Its one- and two-dimensional residual images show that the \textsc{SerSer} model fits (the second row) significantly better than the \textsc{SerExp} model (the first row), effectively describing both the central bulge and the outer disk components. In contrast, galaxy 11867-12701 is disk-dominated and has weak spiral arms. Its radial profile comparison shows that the \textsc{SerExp} model (the third row) provides a better fit. Additionally, when evaluating the two-component decompositions, we observed that although for some galaxies, the \textsc{SerSer} model produces a low overall BIC value, the residuals in the central region are large, indicating a poor fit for the bulge component. Furthermore, galaxies with prominent spiral arms exhibit smaller disk S\'ersic index $n$ and a particularly dim central surface brightness compared to those with weak or absent spiral arms. Therefore, we only adopt the \textsc{SerSer} model when it yields a low BIC value and minimal residuals in the central 20 $\times$ 20 pixel region.
    
    Based on the optimal two-component bulge-disk decompositions, we identified and excluded 105 galaxies best described by a single S\'ersic component. These galaxies typically exhibited bulges with exceptionally high S\'ersic index $n$, abnormally faint or unreliable disk component fits, and large bulge-to-total ratios (B/T). To systematically identify them, we generated histograms of B/T and $\mu_{\rm 0,d,cor}(g)$ (see Section \ref{sec:Selection of LSB galaxies} for calculation details) and visually inspected decomposition results for galaxies in the extreme tails of these distributions. We applied quantitative selection criteria requiring either a B/T ratio greater than 1.5$\sigma$ (i.e., B/T$>$0.66) or a $\mu_{\rm 0,d,cor}(g)$ outside the 2$\sigma$ range (i.e., $\mu_{\rm 0,d,cor}(g) <$ 19.44 mag arcsec$^{-2}$ or $\mu_{\rm 0,d,cor}(g) >$ 24.04 mag arcsec$^{-2}$). After screening out these 105 single-component galaxies, 1,118 galaxies remain for further study.

\begin{figure*}[htbp]%
    \centering
    \includegraphics[width=0.85\textwidth]{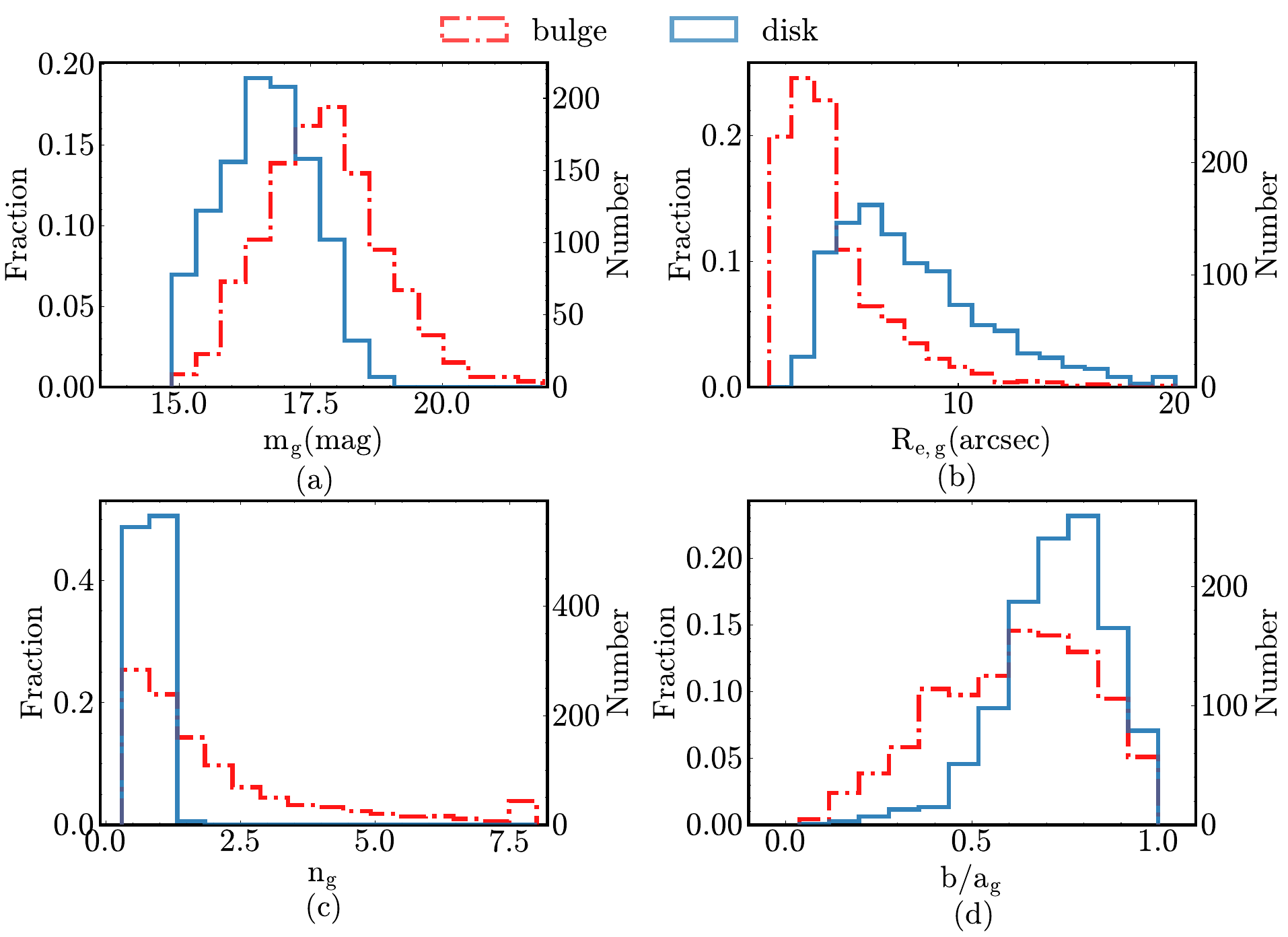}
    \caption{Histogram distributions of the fitted parameters for the bulge (red dashed) and disk (blue solid) components of the sample in the $g$-band: (a) apparent magnitude ($m$); (b) half-light radius ($R_e$) in arcsec; (c) S\'ersic index $n$; (d) axial ratio ($b/a$).}
    \label{fig2}
    \end{figure*}

    The comparison of fitting parameters for the bulge and disk components of these 1,118 galaxies are illustrated in Fig. \ref{fig2}. This figure displays the distribution of integrated magnitude ($m$, see Eq. \ref{formula3}), effective radius ($R_e$), S\'ersic index ($n$) and axial ratio ($b/a$) for the bulge (red dashed lines) and disk (blue solid lines) components in the $g$-band. It shows that, compared to the bulge component, the disk component of our sample has a smaller integrated magnitude, a larger radius, a smaller S\'ersic index $n$, and a slightly larger axial ratio.
    
    \subsection{Selection of LSB galaxies}
    \label{sec:Selection of LSB galaxies}
    
    We selected LSB galaxies based on the central surface brightness of their disk components, as derived above. 
    As described in Section \ref{sec:INTRODUCTION}, we adopt $\mu_{\rm 0,d,cor}(g) \geq$ 22.3 mag arcsec$^{-2}$ as the selection criterion for our LSB galaxy samples. This $g$ band threshold has been calibrated to match the B-band standard (\citealt{zhong_2008}, Eq.3). In this section, we calculate the central surface brightness of the disk component for face-on late-type galaxies (1,118 in total) based on the fitting parameters of the disk components, namely the S\'ersic index ($n_{d}$), effective radius ($R_{e,d}$) and apparent magnitude ($m_{d}$), given by the best-fitting function obtained using GALFIT software.
   %In general, LSB galaxies are defined by the central surface brightness of the disk component in the B-band, where $\mu_{\rm 0,d}(B) \geq$ 22.5 mag arcsec$^{-2}$ \citep{McGaugh_s_1996, Rosenbaum_2009, Du_2015}. Although the B-band central surface brightness can be converted from $g$ and $r$ band values using established color conversion formulas \citep{zhong_2008, Pahwa_2018}, bulge-disk decomposition results are sensitive to the bandpass. Fitting in different bands may yield varying component values, which can impact the robustness of LSB classification. Based on the above considerations, we adopt $\mu_{\rm 0,d}(g) =$ 22 ± 0.3  mag arcsec$^{-2}$ \citep{Tanoglidis_2021} as the selection criterion for our LSB galaxy samples, a value equivalent to that in the B-band. 

    When the radial surface brightness profile of the disk component is characterized by the S\'ersic function \citep{Sersic_1963}, it takes the form:
    \begin{equation}
	I(r)=I_{0}~{\rm exp}\left[-b_{n} \left(\frac{r}{R_{e}}\right) ^{1/n_{d}}\right]
    \label{formula2}
    \end{equation}
    where $I_{0}$ is the central surface brightness of the disk component, and $R_{e}$ denotes the half-light radius of the disk component as specified in the catalog. The parameter $b_{n}$ is defined as $b_{n}= 1.9992 n_{d}-0.3271$ \citep{Graham_2005}, where $n_{d}$ is the S\'ersic index of the disk component obtained from the fit. Notably, when $n=1$ the S\'ersic function reduces to an exponential profile, which is widely used to model the disk component of galaxies. The total luminosity of disk component is obtained by integrating $I(r)$ over the radius, and is directly linked to the fitted disk magnitude m$_{d}$:
    \begin{equation}
        m_{d} = -2.5\log \int_{0}^{\infty} 2\pi r I(r) \rm d \it r
    \label{formula3}
    \end{equation}
    By substituting Eq. \ref{formula2} into the integral, the central surface brightness $I_0$ can be uniquely calculated from the fitted m$_{d}$, given known values of $b_n$ and $R_e$. In logarithmic units, its expression is:
    \begin{equation}
        \mu_{0}=-2.5\log I_0
    \label{formula4}
    \end{equation}
    
    When the disk component is fitted by the exponential function, which corresponds to the special case of the S\'ersic index $n_{d}$=1, the formula for the central surface brightness of the disk can be expressed as \citep{zhong_2008}:
    \begin{equation}
        \mu_{0,d} = m_{d} + 2.5\log (2\pi R_{d}^2)
        \label{formula5}
    \end{equation}
    where $m_{d}$ is the apparent magnitude of the galactic disk, and $R_{d}$ is the scale-length of the disk. For an exponential disk, $R_{d}$= $R_e$/1.678.
    
    We assume that the disk is optically thin, and we corrected its central surface brightness for inclination (using the axial ratio) and cosmic dimming effects, as follows:
    \begin{equation}
        \mu_{\rm 0,d,cor} = \mu_{0,d} + 2.5\log (b/a) - 10\log (1+z)
    \label{formula6}
    \end{equation}
    where $b/a$ is the ratio of the semi-minor axis to the semi-major axis of the disk, and $z$ is the galaxy redshift.
    
\begin{figure}[htbp] 
    \centering
    \includegraphics[width=0.48\textwidth]{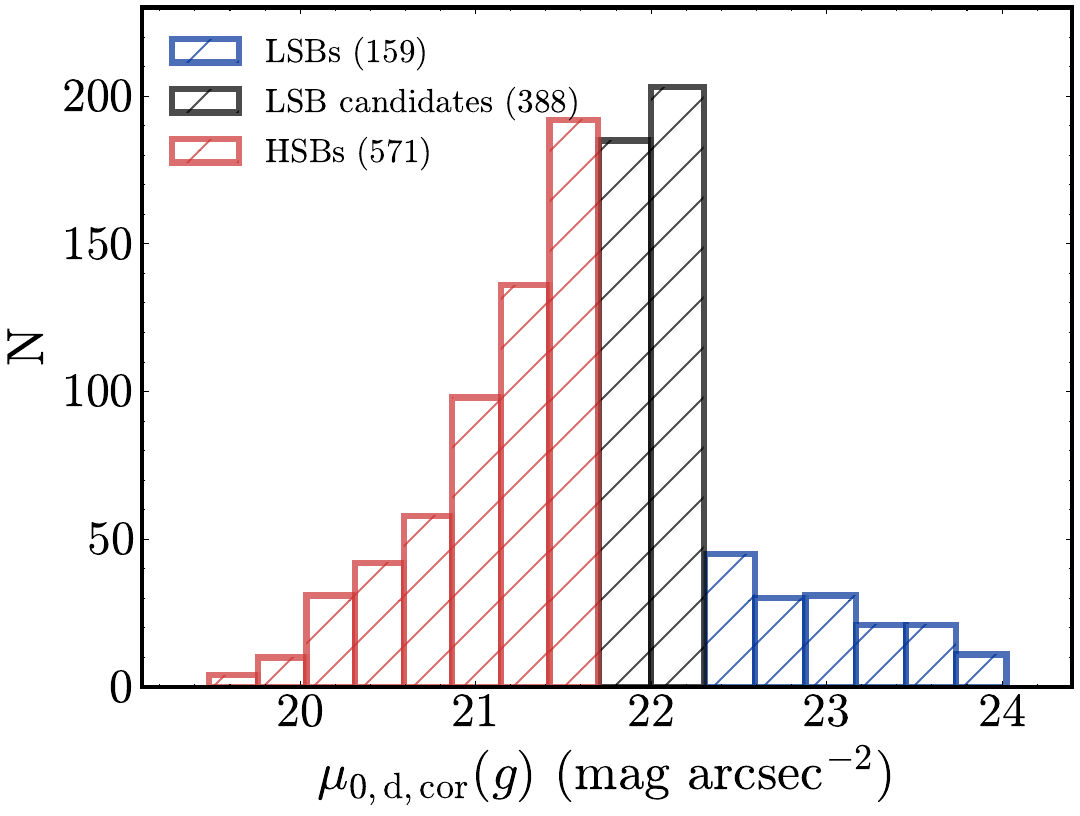}%mu0_disk_b
    \caption{The histogram distribution of the corrected central surface brightness of the disk component in the $g$-band $\mu_{\rm 0,d,cor}(g)$ for the selected face-on late-type galaxies (1,118 in total). The entire sample is divided into three categories: the region where $\mu_{\rm 0,d,cor}(g) \geq$22.3 mag arcsec$^{-2}$ represents LSB galaxies (159, indicated by the blue region), the region with 21.7$< \mu_{\rm 0,d,cor}(g) <$22.3 mag arcsec$^{-2}$ denotes LSB candidates (388, represented by the black region), and the region where $\mu_{\rm 0,d,cor}(g) \leq$21.7 mag arcsec$^{-2}$ as HSB galaxies (571, shown in the red region).}
    \label{fig3}
    \end{figure}

    Using Eq. \ref{formula6}, we calculated the corrected disk central surface brightness of the $g$-band, denoted as $\mu_{\rm 0,d,cor}(g)$, for the entire sample of 1,118 galaxies. The histogram distribution of $\mu_{\rm 0,d,cor}(g)$ is shown in Fig. \ref{fig3}. We classified the galaxies based on the criterion $\mu_{\rm 0,d,cor}(g) =$22$\pm$0.3 mag arcsec$^{-2}$, where the $\pm$0.3 mag arcsec$^{-2}$ uncertainty is consistent with the uncertainty range of the original the central surface brightness of the classical Freeman disk and consider the model dependency of fitted $\mu_{\rm 0,d,cor}(g)$ value. Ultimately, following our previously defined classification criteria (see Section \ref{sec:INTRODUCTION}), we identified 159 LSB galaxies, 388 LSB candidates, and 571 HSB galaxies. In Appendix Fig. \ref{figA1}, we show $g$-band images of all 159 LSB galaxies.
    %The classification is as follows: LSB galaxies ($\mu_{\rm 0,d,cor}(g) \geq$22.3 mag arcsec$^{-2}$), LSB candidates (21.7$< \mu_{\rm 0,d,cor}(g) <$22.3 mag arcsec$^{-2}$), and HSB galaxies ($\mu_{\rm 0,d,cor}(g) \leq$21.7 mag arcsec$^{-2}$).
    
    Total atomic hydrogen masses ($M_{H\,\textsc{i}}$) were obtained from follow-up observations by \citet{Masters_2019}, who detected 144 LSB galaxies, 208 LSB candidates, and 277 HSB galaxies, all of which have H$\,\textsc{i}$ detections. However, the absence of H$\,\textsc{i}$ detection in remaining galaxies does not imply a lack of gas, as some systems fall below the survey's detection limit. The gas depletion time ($t_{\rm dep}$) is calculated using the formula $t_{\rm dep}$ =$M_{\rm gas}$/SFR \citep{McGaugh_2017}, with $M_{\rm gas}$ defined as $M_{H\,\textsc{i}}$ adjusted for cosmic helium abundance $M_{\rm gas}$ =$M_{H\,\textsc{i}}*$1.33 \citep{Eder_2000, Schombert_2011, McGaugh_2017}. The star formation efficiency (SFE) of H$\,\textsc{i}$ gas is calculated using the formula SFE$_{H\,\textsc{i}}$=SFR/$M_{H\,\textsc{i}}$ \citep{Lei_2019}.

    \subsection{Control Sample of LSB and HSB Galaxies}
    \label{sec:Control Sample of LSB and HSB Galaxies}
    
    To investigate the star formation properties of LSB and HSB galaxies while eliminating confounding effects from $M_{\ast}$ and SFR, we compared samples with matched $M_{\ast}$ ($|\Delta \log M_{\ast}| \leq 0.1$ dex) and SFR ($|\Delta \log \rm{SFR}| \leq 0.1$ dex). This ensures the two galaxy populations share comparable fundamental physical properties, allowing for a direct, unbiased comparison of their star formation characteristics. This matching strategy was motivated by key considerations: Firstly, $M_{\ast}$ serves as a fundamental parameter, as many physical properties—such as SFR, metallicities, and stellar populations—vary significantly with stellar mass, making it an ideal comparison metric. Additionally, LSB galaxies are typically characterized by their low star formation rates, providing a unique opportunity to explore how their radial profiles differ under similar SFR conditions as those of HSB galaxies. Ultimately, 155 LSB galaxies were matched with 267 corresponding HSB galaxies, leaving only 4 LSB galaxies unmatched. The matching results are shown in the left panel of Fig. \ref{figA2} of Appendix \ref{The results of the control sample}.
    
    Given the larger population of HSB galaxies, individual LSB galaxies often have multiple HSB counterparts that satisfy the matching criteria. To ensure robust and unbiased statistical comparisons, we applied count weighting to the LSB sample: each LSB galaxy is weighted by the number of its matched HSB counterparts (i.e., an LSB galaxy matched to N HSB galaxies contributes N entries to the statistical analysis). This approach accounts for the sample size imbalance while preserving the relative importance of each LSB galaxy in the comparison. To allow for a better understanding of the trends in star formation and stellar populations, we categorized the matched LSB and HSB galaxies into three stellar mass ranges, namely $\leq$ 10$^{9.5}$ M$_{\odot}$, 10$^{9.5}$-10$^{10}$ M$_{\odot}$, and $\geq$ 10$^{10}$ M$_{\odot}$. This categorization maximizes the number of galaxies in each range to ensure statistical significance.
    
    \subsection{Measurements of SFR and Gas Phase Metallicity}
    \label{sec:Measurements of parameters in MaNGA data}
    
   Following \citet{Catal_2015}, we derive the H$\alpha$ emission line dust attenuation ($A(H\alpha$)) from emission line fluxes, using the H$\alpha$ flux provided by MaNGA spectroscopic data. The formula is as follows:
    \begin{equation}
        \begin{aligned}
            A(H\alpha)=\frac{2.5 \times k(\lambda_{H\alpha})}{k(\lambda_{H\beta})-k(\lambda_{H\alpha})}×\log (\frac{F_{H\alpha}}{2.86 \times F_{H\beta}})
            \label{formula7}
        \end{aligned}
    \end{equation}
    where $k(\lambda$) is the galaxy dust extinction curve, $k(\lambda_{H\alpha}$) = 2.53 and $k(\lambda_{H\beta}$) = 3.61 are the extinction coefficients for the Galactic extinction curve from \citet{Cardelli_1989}. In this section, dust attenuation($A_V$) can be estimated using the dust attenuation of H$\alpha$ emission line. Assuming an extinction curve of $R_V$ =3.1 \citep{Cardelli_1989}, the optical extinction is given by \citep{Barrera_2020}:
    \begin{equation}
        \begin{aligned}
            A_V=A(H\alpha)/0.817
            \label{formula8}
        \end{aligned}
    \end{equation}
    
    The SFR can also be estimated from the H$\alpha$ luminosity. However, since the flux of the H$\alpha$ emission line is affected by dust extinction, we employ Formula \ref{formula7} to perform a dust correction before calculating the SFR. The H$\alpha$ luminosity, after dust extinction correction, is given by the following equations \citep{Calzetti_1994, Calzetti_1997}:
    \begin{equation}
        \begin{aligned}
            L_{H\alpha,\rm int}=L_{H\alpha,\rm obs} \times 10^{0.4 \times A(H\alpha)}
            \label{formula9}
        \end{aligned}
    \end{equation}
	where $L_{H\alpha,\rm obs}$ is the observed H$\alpha$ luminosity.
    
    \citet{Kennicutt_1998} adopted the \citet{Salpeter_1955} initial mass function (IMF) and proposed an expression for estimating the SFR based on H$\alpha$ luminosity \citep{Steffenz_2021, Kennicutt_2012}:
    \begin{equation}
        \begin{aligned}
            \text{SFR}({\rm M_\odot/yr})=7.9 \times 10^{-42} L_{H\alpha}
            \label{formula10}
        \end{aligned}
    \end{equation}
    
    The average SFR surface density ($\Sigma_{\rm SFR}$) characterizes the SFR per unit area, calculated as the total SFR divided by the total area of the galaxy. The spatially resolved $\Sigma_{\rm SFR}$ is estimated within individual spaxel of MaNGA. We calculate the SFR for each spaxel using the dust-extinction-corrected H$\alpha$ luminosity. MaNGA has a spaxel scale of 0.5 arcsec, corresponding to a spixel area of 0.25 arcsec$^2$. This angular scale is converted to physical units in kpc using the galaxy's redshift. The $\Sigma_{\rm SFR}$ for each spaxel is derived by normalizing the spaxel-specific SFR by its physical area. Additionally, to ensure data quality, only spaxels with an H$\alpha$ line signal-to-noise (S/N) $\geq$ 5 are included in the analysis.

    Gas phase metallicity (12+log(O/H)) can be measured from the fluxes of gas emission lines in the galaxies. We use the MaNGA spectroscopic data to apply the analysis outlined by \citet{Tremonti_2004} and use the R23 method to estimate the gas-phase metallicity as follows:
    \begin{equation}
        \begin{aligned}
            12+\log {\rm(O/H)}=9.185-0.313x-0.264x^2-0.321x^3
            \label{formula11}
        \end{aligned}
    \end{equation}
    Where $x$ = log(([O$\,\textsc{ii}$]$\lambda\lambda$3726, 3729 + [O$\,\textsc{iii}$]$\lambda\lambda$4959, 5007)/H$\beta$), with [O$\,\textsc{ii}$] and [O$\,\textsc{iii}$] being the sum of the fluxes of the respective emission lines. 

    We also compare the gas-phase metallicities derived as described above with those determined by the R calibration described in \citeauthor{Pilyugin_2016} (\citeyear{Pilyugin_2016}, hereinafter referred to as PG16) and the N2S2H$\alpha$ calibrator from \citeauthor{Dopita_2016} (\citeyear{Dopita_2016}, hereinafter referred to as DOP16). Both calibrations incorporate the [N$\,\textsc{ii}$]/H$\alpha$ term, which is considered less sensitive to ionization parameters compared to other diagnostic ratios, such as [O$\,\textsc{ii}$]/H$\beta$ and [O$\,\textsc{iii}$]/H$\beta$ in the R23 method \citep{Hwang_2019}. The formula for calculating gas-phase metallicity using the $R$ calibration in PG16 is:
    \begin{equation}
        \begin{aligned}
            12+\log {\rm(O/H)}=a_1+a_2 \log(R_3/R_2)+a_3 \log N_2 \\
            +(a_4+a_5 \log(R_3/R_2)+a_6 \log N_2) \times \log R_2
            \label{formula12}
        \end{aligned}
    \end{equation}
    where $R_2$ = [O$\,\textsc{ii}$]$\lambda \lambda$3727,3729/H$\beta$, $R_3$ = [O$\,\textsc{iii}$]$\lambda \lambda$4959, 5007/H$\beta$ and $N_2$ = [N$\,\textsc{ii}$]$\lambda \lambda$6548,6584/H$\beta$, with [N$\,\textsc{ii}$] indicates that line's flux. For the star forming region (upper branch) where log$N_2 \geq$ −0.6, the coefficients are $a_1$ = 8.589, $a_2$ = 0.022, $a_3$ = 0.399, $a_4$ = −0.137, $a_5$ = 0.164, $a_6$ = 0.589. For the metallicity of the lower branch (log$N_2 <$ −0.6), the coefficients are: $a_1$ = 7.932, $a_2$ = 0.944, $a_3$ = 0.695, $a_4$ = 0.970, $a_5$ = −0.291, $a_6$ = −0.019.
    
    Besides, the N2S2H$\alpha$ calibrator proposed by DOP16 is expressed as:
    \begin{equation}
        \begin{aligned}
            12+\log {\rm(O/H)}=8.77 + y
            \label{formula13}
        \end{aligned}
    \end{equation}
    where $y$ = log[N$\,\textsc{ii}$]$\lambda$6584 / [S$\,\textsc{ii}$]$\lambda \lambda$6717,6731 + 0.264 log[N$\,\textsc{ii}$]$\lambda$6584 / H$\alpha$. Similar to the calculation of $\Sigma_{\rm SFR}$, only spaxels with an S/N $\geq$ 5 for the emission line are considered when calculating the 12+log(O/H) value.

    To investigate the local star formation properties of LSB galaxies as a function of radius, we begin by constructing radial profiles of each parameter, namely: stellar mass surface density ($\Sigma_{\ast}$), $\Sigma_{\rm SFR}$, specific star formation rate (sSFR), 12+ log(O/H), $A_{V}$, $D_{n}$4000, H$\delta_A$, and luminosity/mass-weighted Age. To correct for galaxy's inclination and ensure radial distances correspond to physical radii, we generate several concentric elliptical rings using the galaxy's axial ratio ($b/a$) and major axis position angle ($\phi$). The rings are centered on the galaxy's position (x$_0$, y$_0$) with widths of 0.2 $R_e$, ranging from 0.0 to 1.4 $R_e$. For each ring, we calculate the median value of the physical quantity under consideration, which allows us to derive the radial distribution profile of the galaxy parameters. The MaNGA sample does not have uniform spatial coverage, with 63\% of the data covering up to 1.5 $R_e$, and the remaining 37\% not beyond 2.5 $R_e$ range. This implies there are fewer selected spaxels beyond 1.5 $R_e$ \citep{Steffen_2021}. Consequently, we limit our analysis of radial profiles to a maximum of 1.4 $R_e$.
    
    \section{RESULTS}
    \label{sec:RESULTS}

    In Section \ref{sec:Global Properties}, we analyze the global properties, which consist of the scaling relations between various parameters and $M_\ast$, as well as the correlations between parameters and the corrected disk center surface brightness across different mass ranges. A subsequent analysis of spatially resolved properties will be presented in Section \ref{sec:Radial Profiles and Gradients}.

    \subsection{Global Properties}
    \label{sec:Global Properties}

    \subsubsection{The Relationships of $M_\ast$ with Different Parameters}
    \label{subsubsection:The Relationships of M with Different Parameters}
    
    In section, we analyze key physical parameters and their correlations for our sample of LSB galaxies, with systematic comparisons to HSB galaxies. The parameters examined include SFR, which reflects the star formation properties of the galaxies, gas-phase metallicity (12+log(O/H)), which indicates the enrichment level of the interstellar gas, and effective radius ($R_e$), which describes their size.
    
\begin{figure*}[htbp]%
    \centering
    \includegraphics[width=0.98\textwidth]{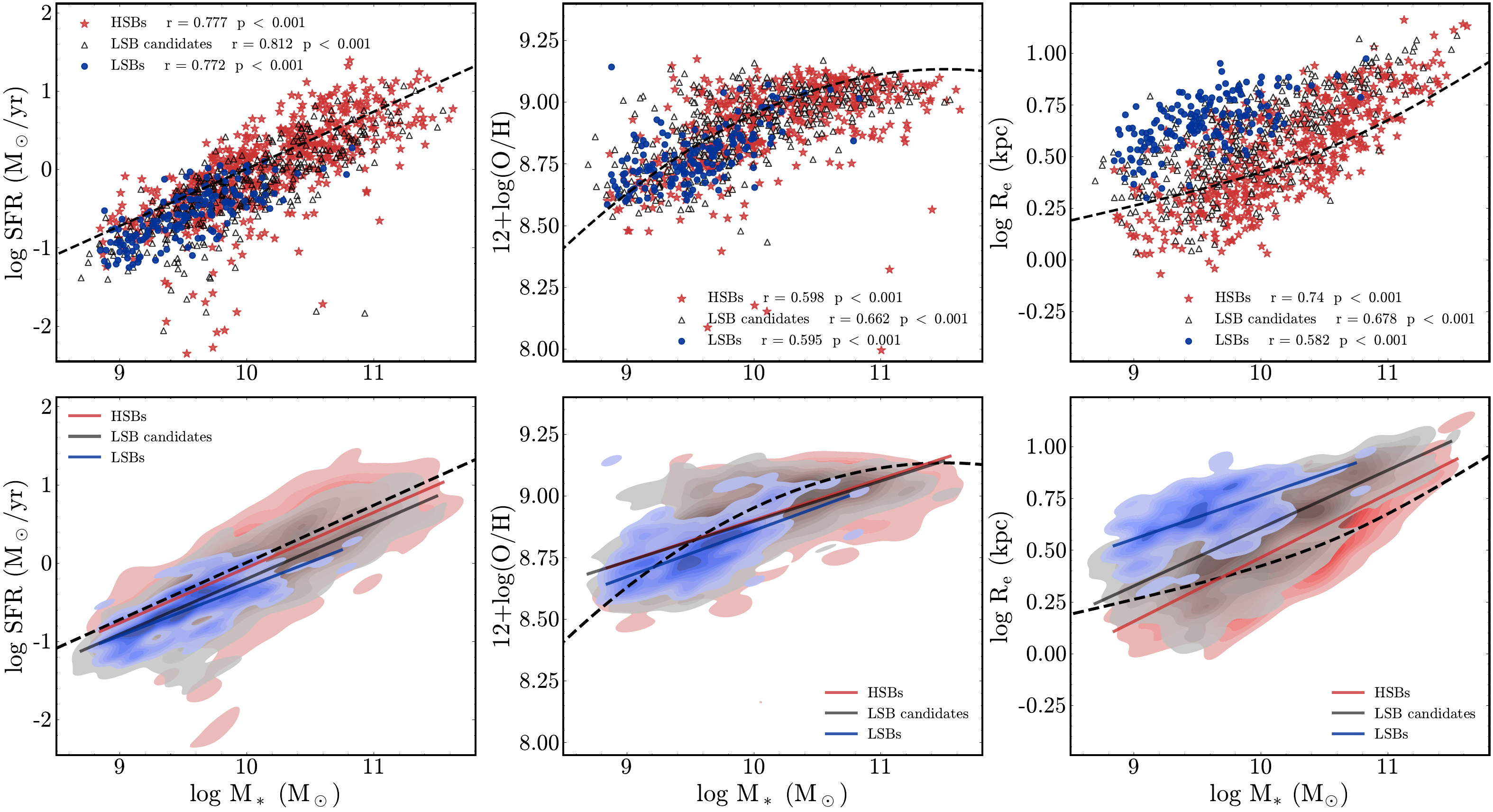}
    \caption{The relationship between stellar mass and SFR, 12+log(O/H) and size. The first row presents a scatter plot, while the second row displays a number density contour. Blue circles and contour represent LSB galaxies, hollow triangles and gray contour denote LSB candidates, while red stars and contour indicate HSB galaxies. In the left panel, the black dashed line depicts the SFMS derived from MaNGA data \citep{Belfiore_2018}. The black dashed line in the median panel illustrates the mass-metallicity relationship for star-forming galaxies presented by \citet{Tremonti_2004}. In the right panel, the black dashed lines represent the mass-size relation for late-type galaxies based on SDSS data \citep{Shen_2003}. In the second row, the blue, gray, and red solid lines represent the quantile regression lines for LSB galaxies, LSB candidates, and HSB galaxies at $q$ = 0.5, respectively. Meanwhile, in the legend of each graph, we also present the $r$ and the $p$ value for each sample provided by Spearman rank correlation.}
    \label{fig4}
    \end{figure*}

    In Fig. \ref{fig4}, we illustrate the distribution of LSB, LSB candidates, and HSB galaxies across three diagrams: SFR-$M_\ast$, 12+log(O/H)-$M_\ast$, and $R_e$-$M_\ast$. The upper row shows scatter plots highlighting correlations among the sample data. In this row, the blue circles, hollow triangles and red stars represent LSB, LSB candidates and HSB galaxies, respectively. We employed Spearman rank correlation analysis, with the Pearson correlation coefficient ($r$) and the corresponding $p$-values, which quantify the correlation between parameters, are listed in the legend. The lower row displays density contour plots with ten contour levels, representing ten isodensity contours across the estimated probability density range. The blue, black, and red contours for LSB, LSB candidates, and HSB galaxies, respectively. Median linear fits for each sample are overplotted in matching colors. These fits were obtained via Quantile Regression with $q$=0.5, ensuring a robust analysis of the underlying relationships. 
    
    In the SFR-$M_\ast$ plane (left panel of Fig. \ref{fig4}), the black dashed line represents the star formation main sequence (SFMS: log(SFR) = 0.73 $\times$ log($M_\ast$) - 7.29) from MaNGA data \citep{Belfiore_2018}, distinguishing star-forming galaxies from their non-star-forming counterparts. The figure reveals that LSB galaxies primarily occupy the low to intermediate mass range, with a notable scarcity of galaxies possessing stellar masses greater than 10$^{11}$ M$_\odot$. Only three LSB galaxies exceed a stellar mass of 10$^{10.5}$ M$_\odot$. Both LSB and HSB galaxies show strong positive correlations in the SFR–$M_\ast$ relation, with $r \approx 0.77$ ($p < 0.001$), confirming statistically significant trends for both populations. However, at fixed $M_\ast$, LSB galaxies show a 15\% lower SFR than HSB galaxies. Furthermore, LSB galaxies (blue line) are more significantly offset from the SFMS than HSB galaxies (red line), whose fitting line of HSB galaxies aligns more closely with the SFMS.
    
    The gas-phase metallicity of LSB galaxies is generally lower than that of HSB galaxies, as illustrated in the middle panel of Fig. \ref{fig4}. A positive $r$ indicates that a larger stellar mass correlates with higher gas-phase metallicity; this correlation is slightly stronger for HSB galaxies. The relationship between gas-phase metallicity and stellar mass for star-forming galaxies, identified by \citet{Tremonti_2004}, is depicted by the black dashed line and can be expressed as: 12+log(O/H) =−1.492 + 1.847(log $M_\ast$) - 0.08026(log $M_\ast$)$^2$. This trend aligns with expectations based on LSB galaxies' low stellar mass and large size, reinforcing the concept that low-mass galaxies typically show low metallicities. Most LSB galaxies fall below this standard, being 12\% lower gas-phase metallicity than HSB galaxies, highlighting their relatively low value.
    
    Deviations from the $M_\ast$-size relation are anticipated for LSB galaxies. The right panel of Fig. \ref{fig4} depicts the $M_\ast$-size relationship, with the black dashed line illustrating the correlation for late-type galaxies from SDSS without distinguishing between LSB and HSB galaxies, as reported by \citet{Shen_2003}. Both LSB and HSB galaxies display a positive $r$, indicating that larger stellar masses are associated with larger sizes, though this correlation is slightly stronger for HSB galaxies. Importantly, at fixed stellar mass, the effective radius of LSB galaxies is 29\% larger than that of HSB counterparts, with all LSB galaxies located above the dashed line.
    
    We performed an $M_\ast$-SFR matching analysis (see Fig. \ref{figA2} in Appendix \ref{The results of the control sample}), which confirms that even after matching $M_\ast$ and SFR, LSB galaxies still exhibit systematically lower gas-phase metallicities, lower SFRs, and larger $R_e$ compared to HSB galaxies, consistent with our earlier results.
    
    Therefore, in general, compared to HSB galaxies, LSB galaxies generally show lower SFR and 12+log(O/H), while demonstrating larger $R_e$.
    
    \subsubsection{Global Properties for Different Mass Ranges}
    \label{sec:Global Properties for Different Mass Ranges}
    
    To further examine the comparison of global properties between LSB and HSB galaxies across different mass ranges, we categorize each sample into three mass ranges, namely $\leq$ 10$^{9.5}$ M$_{\odot}$, 10$^{9.5}$-10$^{10}$ M$_{\odot}$, and $\geq$ 10$^{10}$ M$_{\odot}$. Each parameter, such as $\Delta$log(SFR), $\Delta$12+log (O/H), and $\Delta$log($R_e$) represent the residuals from established scaling relations mentioned in Section \ref{subsubsection:The Relationships of M with Different Parameters}, including the SFMS, the 12+log(O/H) - $M_\ast$ relation for star-forming galaxies, and the $R_e$ - $M_\ast$ relation for late-type galaxies. By using residuals, this approach removes the primary dependence on stellar mass, allowing a fair comparison between LSB and HSB galaxies relative to standard benchmarks that describe the typical links between stellar mass, star formation, chemical enrichment, and galaxy size in the local Universe \citep{Belfiore_2018,Tremonti_2004,Shen_2003}.

\begin{figure*}[htbp]%
    \centering
    \includegraphics[width=0.82\textwidth]{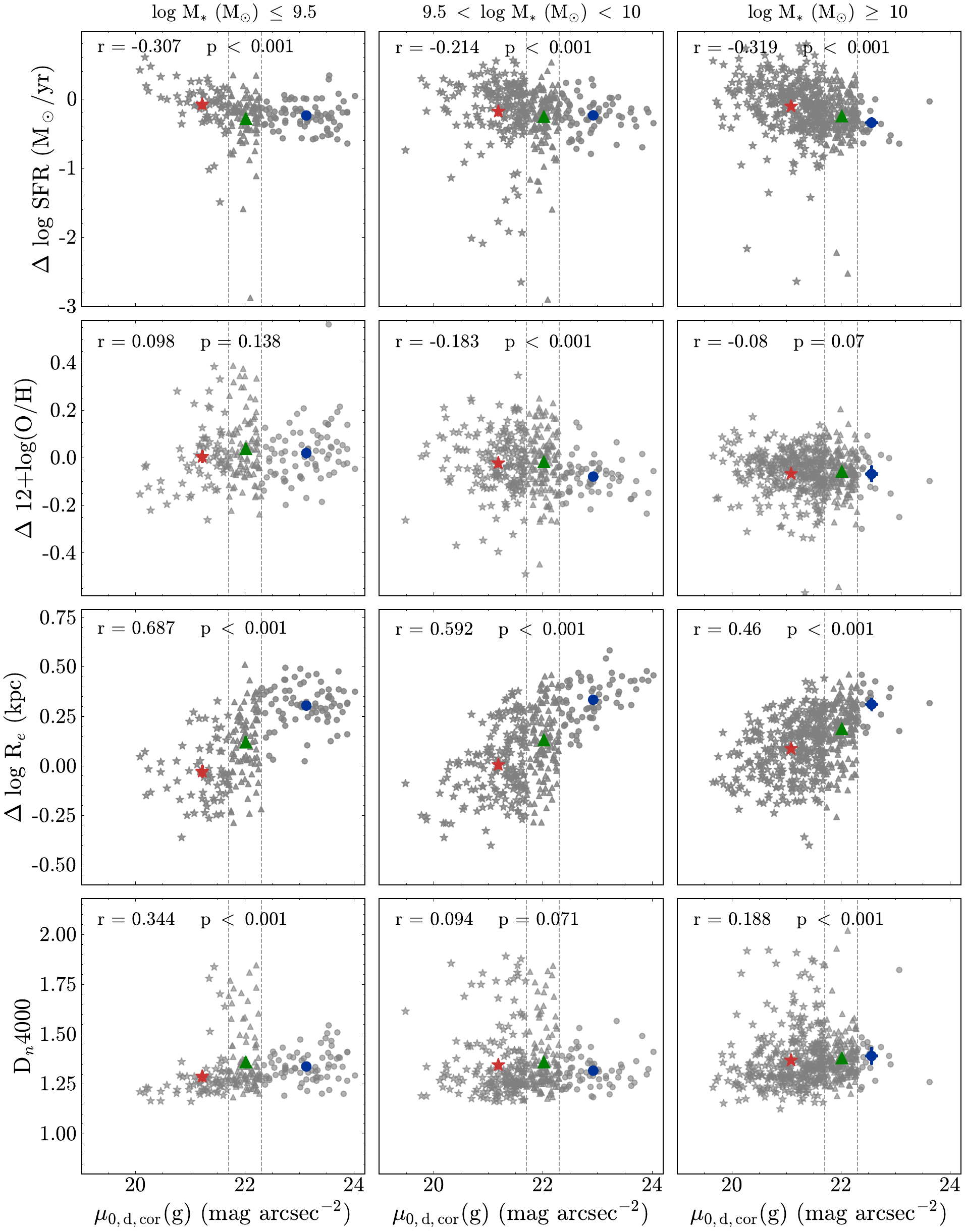}
    \caption{The residuals of key galaxy parameters (SFR: $\Delta$log(SFR); gas-phase metallicity: $\Delta$12+log(O/H); effective radius: $\Delta$log($R_e$))-calculated relative to correlations from reference galaxy samples-along with D$_n$4000 values, plotted against the corrected disk central surface brightness ($\mu_{\rm 0,d,cor}$(g)) across three stellar mass ranges. The first, second and third columns correspond to galaxies with $M_\ast \leq 10^{9.5}$ M$_\odot$, $10^{9.5} <$ $M_\ast < 10^{10}$ M$_\odot$, and $M_\ast \geq 10^{10}$ M$_\odot$, respectively. Gray points (in distinct shapes) denote individual galaxies in this study: circles for LSB galaxies, triangles for LSB candidates, and stars for HSB galaxies. Larger colored symbols (blue circles for LSB galaxies, green triangles for LSB candidates, red stars for HSB galaxies) represent the weighted averages of each sample. Error bars indicate median measurement uncertainties, while gray dashed lines mark central disk surface brightness boundaries separating HSB, LSB candidates, and LSB galaxies (from left to right). The Pearson correlation coefficients ($r$) and $p$-values for all galaxies in each mass range are displayed in the upper left corner of each panel.}
    \label{fig5}
    \end{figure*}

\begin{figure*}[htbp]%
        \centering
        \includegraphics[width=0.85\textwidth]{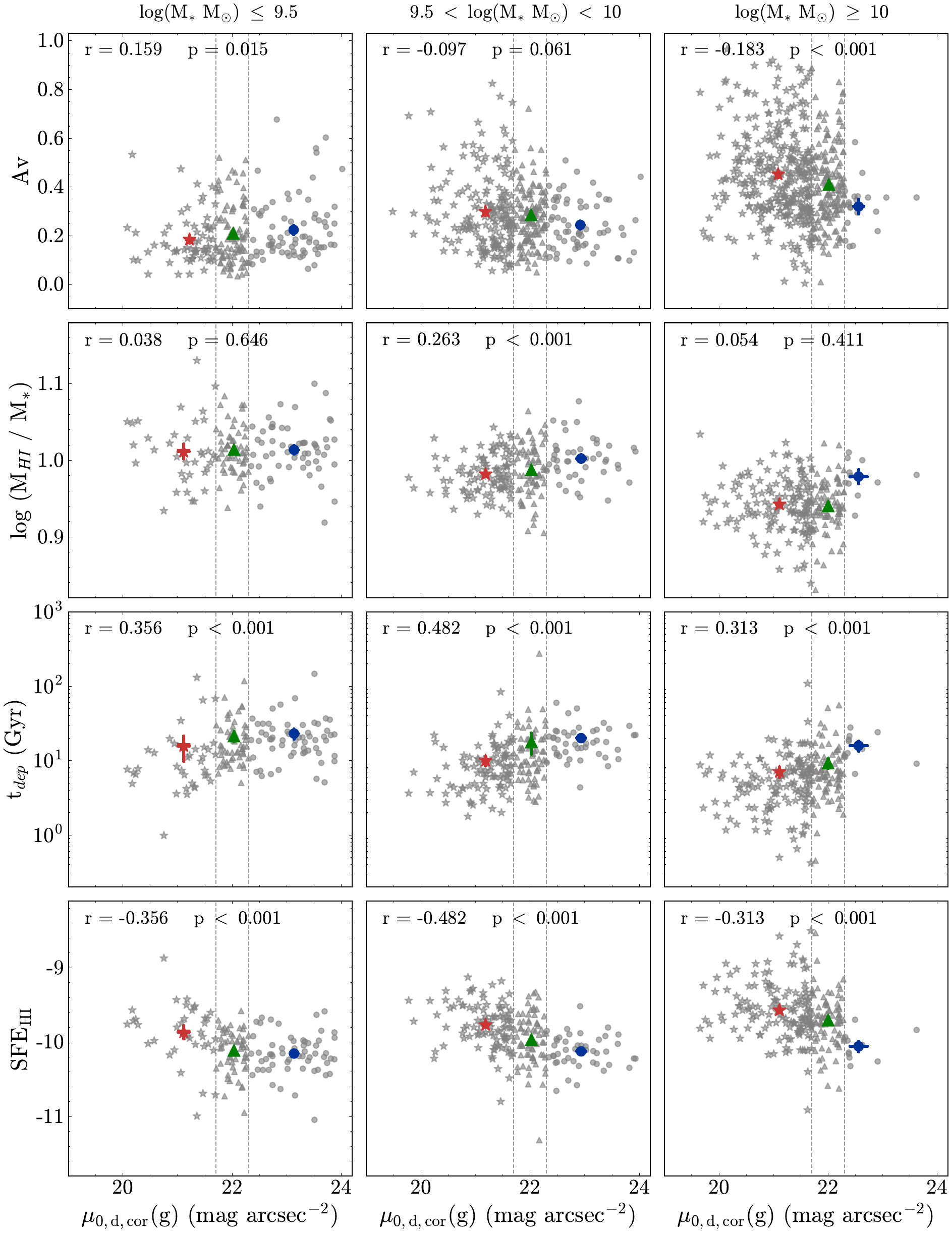}
        \caption{Same as Fig. \ref{fig5}, this figure presents four additional galaxy properties: dust attenuation in the rest-frame V band ($A_{V}$)，the H$\,\textsc{i}$-to-stellar mass ratio ($M_{H\,\textsc{i}}$/$M_\ast$), the gas depletion time ($t_{\rm dep}$) and the star formation efficiency (SFE).}
        \label{fig6}
    \end{figure*}

    Fig. \ref{fig5} and \ref{fig6} illustrate the relationships between $\Delta$log(SFR), $\Delta$12+log(O/H), $\Delta$log($R_e$), $D_n$4000, $A_{\rm V}$, log($M_{\rm H\,\textsc{i}}/ M_\ast$), $t_{\rm dep}$ and SFE$_{\rm H\,\textsc{i}}$ with the corrected central surface brightness of the disk component ($\mu_{\rm 0,d,cor}$). In these figures, the left side of the dashed line denotes HSB galaxies, the middle segment corresponds to LSB candidates, and the right side represents LSB galaxies. Individual galaxies are shown in grey—stars for HSB galaxies, triangles for LSB candidates, and circles for LSB galaxies. The colored markers represent the median values for each population (red for HSB galaxies, green for LSB candidates, and blue for LSB galaxies), with error bars denoting the median uncertainties. These errors are estimated by calculating the sample standard deviation ($\sigma$) and dividing by $\sqrt{n}$, where n is the number of samples. The Pearson correlation coefficient ($r$) and the corresponding $p$-value are shown in the upper left corner of each panel.
    
    In summary, Fig. \ref{fig5} and \ref{fig6} reveal statistically significant ($p <$ 0.001–0.05) but generally weak correlations with disk central surface brightness. The strongest correlation is observed for effective radius with a correlation coefficient $r =$ 0.5–0.7 that varies with $M_\ast$. Compared to HSB galaxies, LSB galaxies show lower $\Delta \log(\mathrm{SFR})$ and $\Delta 12 + \log(\mathrm{O/H})$, and significantly larger $\Delta \log(R_e)$, with these offsets increasing toward higher stellar masses. This indicates a systematic deviation from standard galaxy scaling relations, particularly for massive LSB galaxies, and suggests a distinct evolutionary pathway likely associated with high angular momentum halos and suppression of star formation. LSB galaxies also show slightly higher $D_n$4000 and lower dust extinction. For similar H$\,\textsc{i}$ masses, LSB galaxies have longer gas depletion times than HSB galaxies, indicating lower star formation efficiency and possibly more diffuse H$\,\textsc{i}$ distributions. However, these trends may have been diluted by averaging over large apertures, motivating the need for a spatially resolved analysis, which we present next.
    
    \subsection{Radial Profiles and Gradients}
    \label{sec:Radial Profiles and Gradients}

    Drawn upon the integral field spectroscopic data from the MaNGA survey, we discuss in this section the radial profiles and gradients of key parameters such as $\Sigma_{\ast}$, $\Sigma_{\rm SFR}$, sSFR (sSFR=$\Sigma_{\rm SFR}$/$\Sigma_{\ast}$), 12+ log(O/H), $A_{V}$, $D_{n}$4000, H$\delta_A$, and luminosity/mass-weighted Age in different stellar mass ranges (i.e., $\leq$ 10$^{9.5}$ M$_{\odot}$, 10$^{9.5}$-10$^{10}$ M$_{\odot}$, and $\geq$ 10$^{10}$ M$_{\odot}$). The radial profiles of individual galaxy parameters are presented in Appendix \ref{Individual galaxies, individual bulge components and median profiles in different mass ranges}, where we compare these findings to those of HSB galaxies. This comparison aims to deepen our understanding of the internal and external structural characteristics of LSB galaxies, thereby elucidating their evolutionary mechanisms and associated physical processes. %We also matched each LSB galaxy with a corresponding HSB counterpart by matching} for $M_\ast$ and SFR, as detailed in Section \ref{sec:Control Sample of LSB and HSB Galaxies}. The radial profiles of these matched pairs can be found in Fig. \ref{figA3} of the Appendix \ref{The results of the control sample}.}
    
    We find that LSB galaxies in our sample systematically show larger sizes than HSB galaxies at similar stellar masses (see Section \ref{subsubsection:The Relationships of M with Different Parameters} and the rightmost panel of Fig. \ref{fig4}). This size difference may have a substantial impact on the radial profiles of many physical parameters when comparing LSB galaxies with HSB galaxies. As shown in \citet{Lin_2024}, galaxies with larger sizes typically show lower $\Sigma_\ast$, $\Sigma_{\rm SFR}$, sSFR and 12+log(O/H). Consequently, some of the observed differences in the radial profiles of LSB galaxies comparing with HSB galaxies may reflect the size effects rather than being directly driven by central surface brightnesses. To investigate the effect of galaxy sizes on the comparisons of LSB and HSB galaxies, we introduce an additional comparative sample from \citet{Lin_2024}: a sample of large-size star-forming galaxies selected purely based on their sizes, independent of their central surface brightnesses. These galaxies are drawn from the MaNGA survey and are required to satisfy two criteria: first, they have to be star-forming galaxies by having $|\Delta$ log(SFR)$| < 0.5$ and $>50\%$ star-forming spaxels per galaxy. Then we select those that are located above the black dotted line in the mass–size relation shown in the rightmost panel of Fig.~\ref{fig4}, and define them as the large-size star-forming galaxies. This large-size sample is further divided into two stellar-mass bins: a low-mass bin at 10$^{9}$–10$^{9.5}$ M$\odot$ and an intermediate-mass bin at 10$^{9.5}$–10$^{10}$ M$\odot$. The sample can include both LSB and HSB galaxies, provided that they are classified as having larger-than-average sizes at a given stellar mass. By comparing LSB galaxies with this large-size sample—which shares their large-size property but not necessarily their low surface-brightness characteristics—we may identify the specific role that galaxy size plays in shaping the observed radial profiles.

    Notably, given the uncertainty in the measurements of central surface brightness for our LSB candidates, we do not explicitly incorporate this sample into the subsequent analysis. Instead, we restrict our comparisons to three well-defined samples: LSB galaxies, HSB galaxies, and large-size star-forming galaxies. This approach simplifies the analysis and allows for a clearer interpretation of the results.
    
    \subsubsection{Radial Profiles of Stellar Mass}
    \label{sec:Radial Profiles of Stellar Mass}
    
\begin{figure*}[htbp]%
    \centering
    \includegraphics[width=0.9\textwidth,height=0.32\textheight]{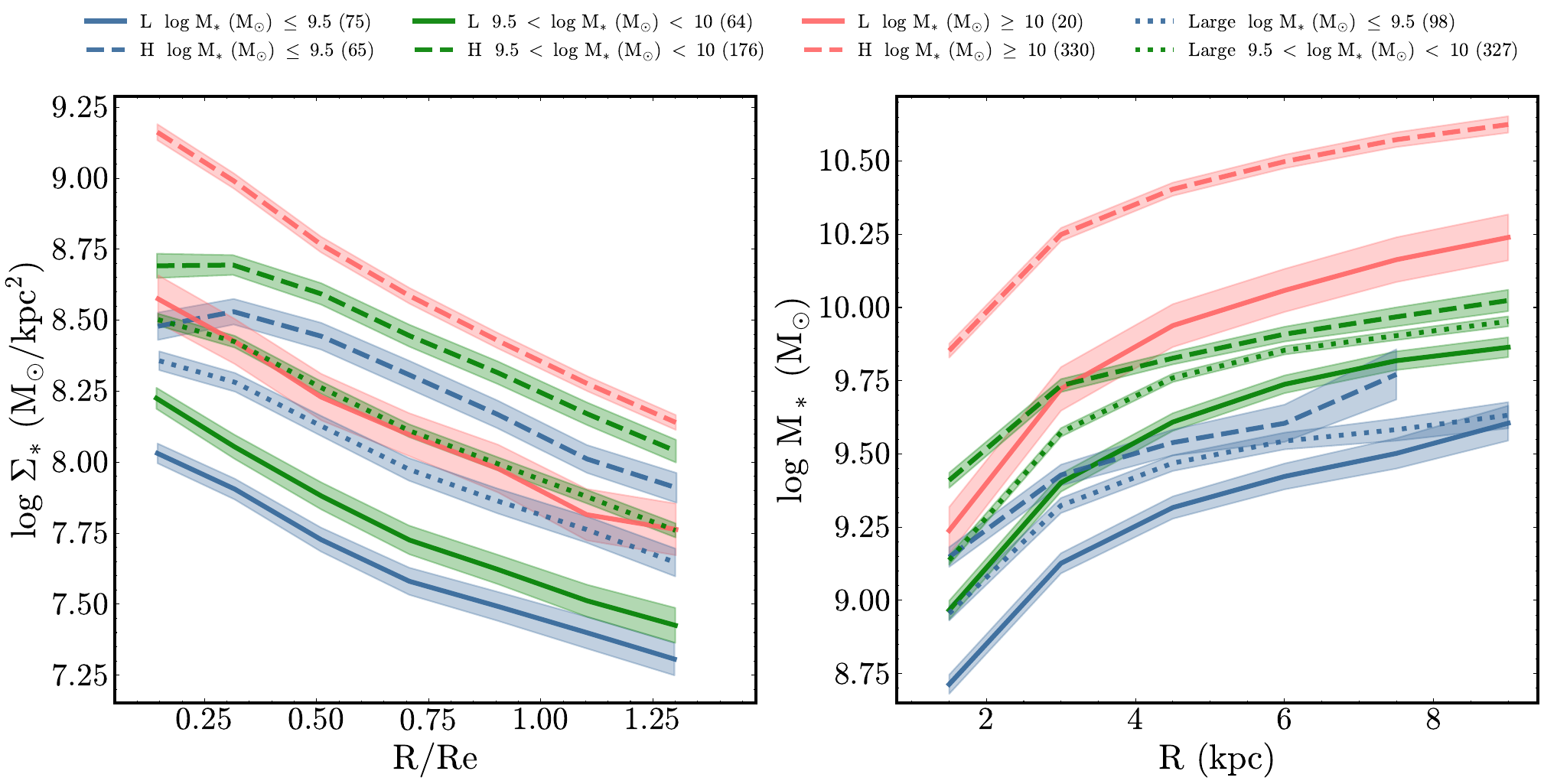}
    \caption{The radial profile of $\Sigma_{\ast}$ as a function of $R$/$R_e$ and the cumulative $M_\ast$ as a function of physical radius $R$ in different stellar mass ranges. Blue, green, and red correspond to the mass ranges: $\leq$ 10$^{9.5}$ M$_{\odot}$, 10$^{9.5}$ -10$^{10}$ M$_{\odot}$, and $\ge$ 10$^{10}$ M$_{\odot}$, respectively. Within each mass range, the solid line represents the median profile of LSB galaxies, the dashed line for HSB galaxies, and the dotted line for large-size galaxies. Shaded areas denote the error associated with each median profile. The number of LSB and HSB galaxies in each stellar mass range is listed in the legend.}
    \label{fig7}
    \end{figure*}

    First, we compare the stellar mass distribution ($M_\ast$) of LSB and HSB galaxies. In Fig. \ref{fig7}, the left panel displays radial profiles of $\Sigma_{\ast}$ as a function of normalized radius ($R/R_e$) for LSB and HSB galaxies across different stellar mass ranges, as well as for large-size galaxies in the low- and intermediate-mass regimes. The right panel, in contrast, shows the cumulative stellar mass $M_\ast$($<R$) plotted against galactocentric radius ($R$). Colored lines in the figure represent the mass-weighted median trends, with the color gradient from blue to red indicating increasing stellar mass. Solid lines correspond to LSB galaxies, dashed lines to HSB galaxies, and dotted lines to large-size star-forming galaxies.

    From the radial profiles of $\Sigma_{\ast}$ shown in the left panel of Fig. \ref{fig7}, it is evident that $\Sigma_{\ast}$ increases systematically with stellar mass, as indicated by the color transitioning from blue (low-mass range) to red (high-mass range). Across all stellar-mass ranges, LSB galaxies consistently exhibit lower $\Sigma_{\ast}$ than their HSB counterparts. This is expected, given that LSB galaxies are typically more extended than HSB galaxies, which naturally leads to lower stellar-mass surface densities. However, it is noteworthy that the radial gradients of $\Sigma_{\ast}$ in LSB galaxies closely resemble those of HSB galaxies, suggesting that the suppression of $\Sigma_{\ast}$ in LSB galaxies occurs uniformly across all $R/R_e$, rather than being confined to either the central or outer regions. Comparatively, the large-size galaxy sample (dotted lines) show lower $\Sigma_{\ast}$ than HSB galaxies, but noticeably higher $\Sigma_{\ast}$ than LSB galaxies across the full range of $R/R_e$. This result indicates that even at comparable sizes, galaxies with lower central surface brightness systematically possess lower $\Sigma_{\ast}$ at all radii.
    
    The right panel of Fig. \ref{fig7} shows the cumulative stellar mass profile, $M_\ast(< R)$, as a function of radius $R$, with stellar mass ranges color-coded in the same manner as in the left panel. $M_\ast(< R)$ increases monotonically with radius for all samples; however, LSB galaxies exhibit a noticeably slower growth of enclosed stellar mass with radius compared to HSB galaxies, and to a lesser extent, the large-size galaxy sample. This behavior indicates that, on an absolute radial scale, the stellar mass distribution of LSB galaxies is the most spatially extended among the three samples, corresponding to a more centrally depleted mass profile relative to both HSB and large-size galaxies.
    
    \subsubsection{Radial Profiles and Gradients of Star Formation}
    \label{sec:Radial Profiles and Gradients of Star Formation}
    
\begin{figure*}[htbp]%
    \centering
    \includegraphics[width=0.8\textwidth,height=0.55\textheight]{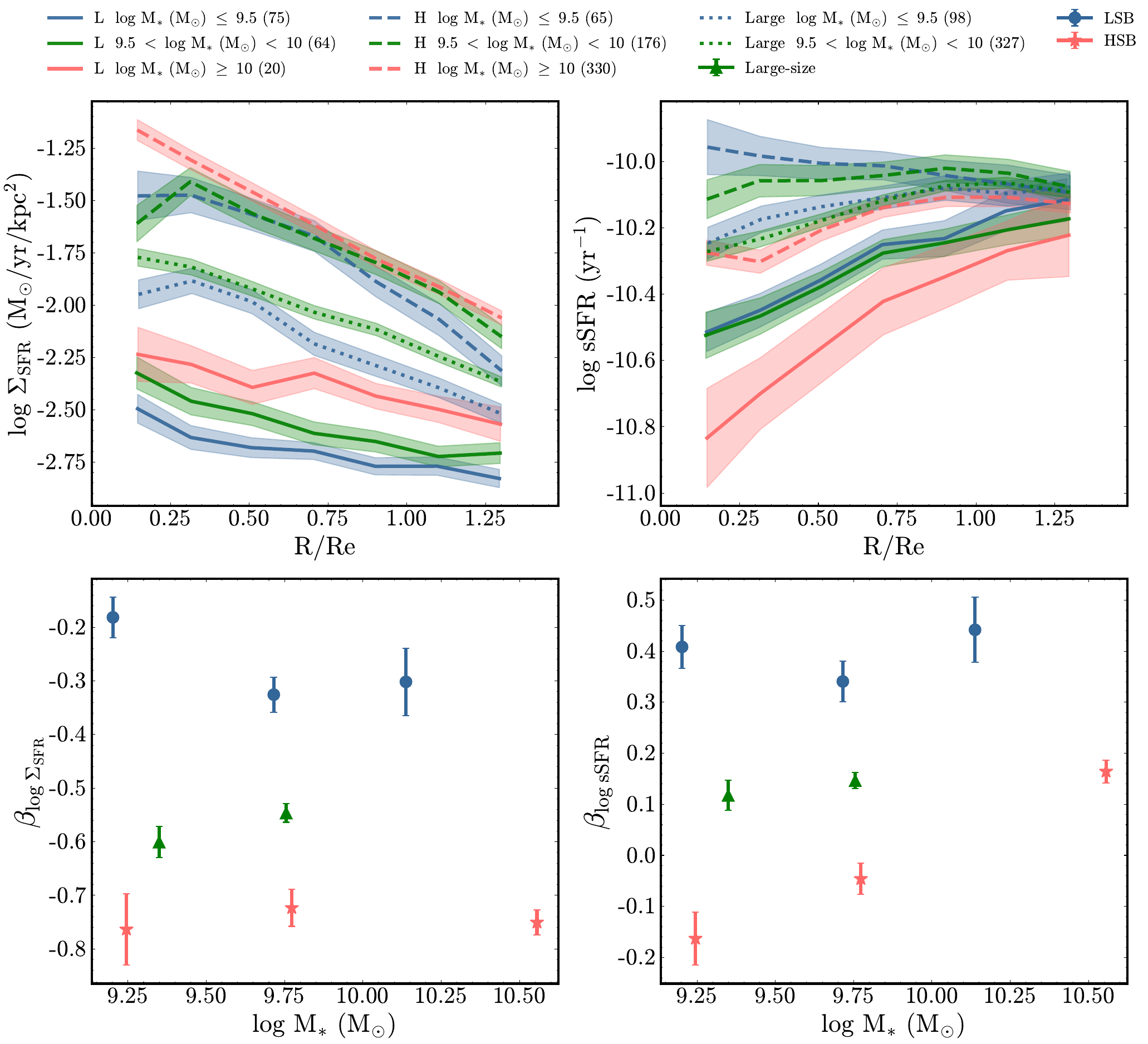}
    \caption{Top row: radial profiles of $\Sigma_{\rm SFR}$ and sSFR across different stellar mass ranges. The lines in the figure correspond to those defined in Fig. \ref{fig7}. Bottom row: relationships between the median radial gradients of $\Sigma_{\rm SFR}$ and sSFR with stellar mass. LSB, HSB and large-size galaxies are represented by blue circles, red stars, and green triangles, respectively, with their associated error bars.}
    \label{fig8}
    \end{figure*}

    To investigate the star formation characteristics of LSB galaxies, we conducted a detailed analysis of their $\Sigma_{\rm SFR}$ and sSFR. The first row of Fig. \ref{fig8} illustrates the radial profiles of $\Sigma_{\rm SFR}$ and sSFR for LSB, HSB, and large-size galaxies in different mass ranges. The color coding of lines in this figure follows the same scheme as in Fig. \ref{fig7}. For our analysis, we computed radial gradients $\beta$ for each galaxy by fitting linear models, $y = \alpha + \beta x$, to their radial profiles, where x represents the radius in units of $R$/$R_e$. The median gradient $\beta$ values within different mass ranges are shown in the second row of Fig. \ref{fig8}. In this plot, blue circles represent LSB galaxies, red stars indicate HSB galaxies, and green triangles denote large-size galaxies.
    
    The first column of Fig. \ref{fig8}, which presents the radial profile and gradient of $\Sigma_{\rm SFR}$, shows that $\Sigma_{\rm SFR}$ gradually increases with stellar mass. Importantly, the $\Sigma_{\rm SFR}$ of LSB galaxies in different mass ranges consistently remain lower than that of HSB galaxies, which is consistent with the observed global properties of the SFMS. When examining the radial gradients, we find minimal variation across different mass ranges. Both LSB and HSB galaxies show negative gradients, but LSB galaxies display significantly flatter gradients. This suggests a smaller difference in $\Sigma_{\rm SFR}$ between the centers and outskirts of LSB galaxies compared to HSB galaxies. Furthermore, when comparing the $\Sigma_{\rm SFR}$ of LSB galaxies with large-size galaxies, both types show lower $\Sigma_{\rm SFR}$ and flatter gradients, yet LSB galaxies have even lower $\Sigma_{\rm SFR}$ and a flatter gradient overall. These observations suggest a more uniform or slower cessation of star formation across the disk of LSB galaxies. Specifically, accounting for $M_\ast$, SFR, and size, reveals that the central $\Sigma_{\rm SFR}$ of LSB galaxies is notably lower than that of HSB galaxies, while their outskirts are similar. This primarily indicates that LSB galaxies have significantly lower $\Sigma_{\rm SFR}$ at their centers compared to HSB galaxies.
    
    We further calculated the sSFR of galaxies using the formula sSFR=$\Sigma_{\rm SFR}$/$\Sigma_{\ast}$. A higher sSFR value indicates a greater star formation rate per unit of stellar mass. The second column of Fig. \ref{fig8} shows the radial profile and gradient of sSFR. Our observations reveal that sSFR gradually decreases with increasing stellar mass, with LSB galaxies exhibiting significantly lower sSFR values than HSB galaxies in different mass ranges. Regarding the radial gradient, both LSB and HSB galaxies show an increasing gradient as stellar mass rises. Notably, LSB galaxies show a positive gradient, while HSB galaxies tend to display negative or flat gradients. This contrast indicates that LSB galaxies have lower sSFR in their central regions and higher sSFR in their outer regions, whereas HSB galaxies typically have higher sSFR in their centers and lower sSFR in the outskirts. Large-size galaxies also show a trend of low sSFR in their centers and higher sSFR in their outskirt; however, there is a noticeable decline beyond 1 R/$R_e$. Compared to large-size galaxies, LSB galaxies present significantly lower sSFR, particularly in their central regions, while showing a gradient that gradually flattens from the inside out. This pattern indicates potential recent star formation activity in the outskirts of LSB galaxies.
    
\begin{figure}[htbp] 
    \centering
    \includegraphics[width=0.45\textwidth]{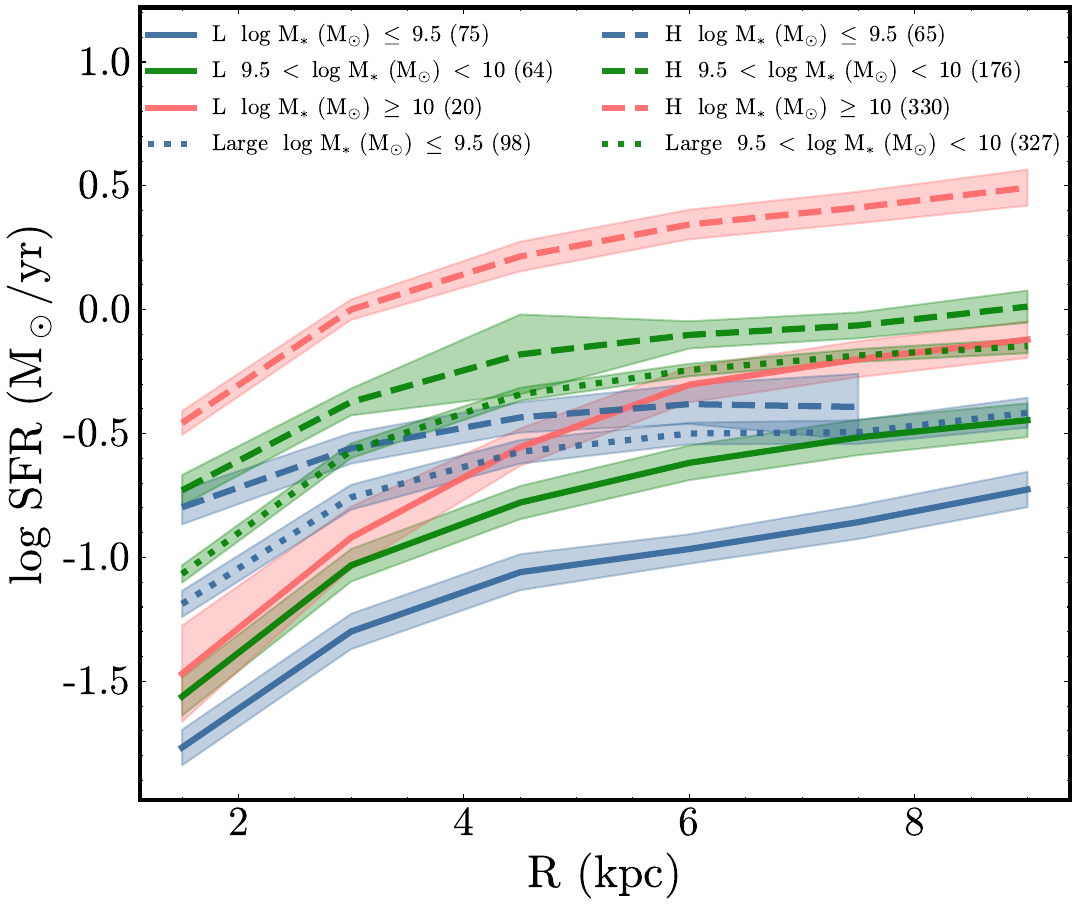}%mu0_disk_b
    \caption{The cumulative SFR as a function of radius $R$, shown across different stellar mass ranges. The lines styles in the figure correspond to those defined in Fig. \ref{fig7}.}
    \label{fig9}
    \end{figure}

    LSB galaxies show significantly lower galaxy-wide SFRs, and their larger sizes result in reduced $\Sigma_{\rm SFR}$. To further investigate the spatial distribution of star formation, we analyze the cumulative SFR within radius R as a function of radius (Fig. \ref{fig9}). The figure reveals that, in different mass ranges, the cumulative SFR at the centers of LSB galaxies is significantly lower than that observed in HSB galaxies. Moreover, the cumulative SFR in HSB galaxies rises steeply and saturates at smaller radii than in LSB galaxies, which reach their maximum cumulative SFR only at considerably larger radius. However, due to notably low central SFR in LSB galaxies, their peripheral and overall cumulative SFR levels remain relatively subdued. Large-size galaxies exhibit a similar trend, whereas their relatively higher central SFR results in a global SFR that is consistently higher than that of LSB galaxies.
    
    In summary, the lower $\Sigma_{\rm SFR}$ observed in LSB galaxies, compared to HSB and large-size galaxies, can be attributed to their larger size. Furthermore, the gradient of $\Sigma_{\rm SFR}$ in LSB galaxies is flatter, and their sSFR is also lower, showing a significantly different positive gradient compared to HSB and large-size galaxies. This behavior may suggest that new star formation activity occurs primarily in the outskirts of LSB galaxies, creating a pattern where star formation rates are reduced in the central regions and enhanced in the outer regions. Furthermore, the central SFR in LSB galaxies is significantly lower than that of HSB and large-size galaxies, leading to reduced global SFRs for LSB galaxies.
    
    \subsubsection{Radial Profile and Gradients of Gas-Phase Metallicity}
    \label{sec:Radial Profile and Gradients of Gas-Phase Metallicity}
    
\begin{figure*}[htbp]%
    \centering
    \includegraphics[width=0.9\textwidth,height=0.32\textheight]{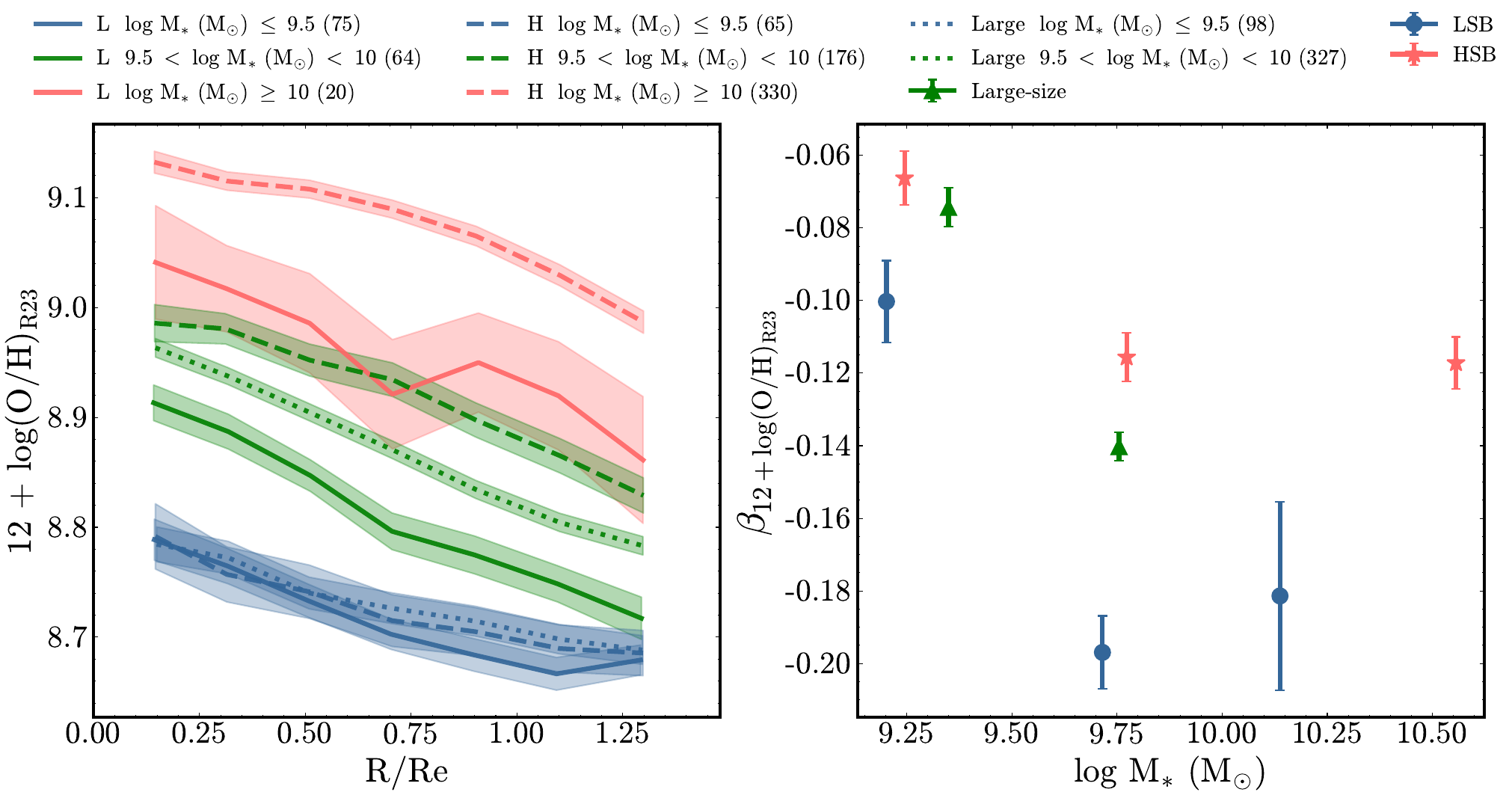}
    \caption{Radial profiles and median gradients of 12+log(O/H)$_{\rm R23}$ for different stellar mass ranges. The color identification of different mass ranges is the same as in Fig. \ref{fig8}.}
    \label{fig10}
    \end{figure*}

    The enrichment history of galaxies is intricately related to their formation and evolution. Gas-phase metallicity serves as a vital indicator of this relationship, directly tracing current enrichment states shaped by star formation and gas accretion \citep{Bao_2021}. Fig. \ref{fig10} presents the radial profile and gradient of 12+log(O/H)$_{\rm R23}$, estimated using the R23 method (described in Section \ref{sec:Measurements of parameters in MaNGA data}) in different mass ranges. In Fig. \ref{figA6} of Appendix \ref{Individual galaxies, individual bulge components and median profiles in different mass ranges}, we further show the gas-phase metallicity derived from the PG16 and DOP16 methods. These two methods provide independent validation and systematic error assessment for our metallicity analysis. 
    
    Analysis of Fig. \ref{fig10} and Fig. \ref{figA6} reveals that as the stellar mass increases, the 12+log(O/H)$_{\rm R23}$ gradually rises. Only in the intermediate and high-mass range do LSB galaxies show lower 12+log(O/H)$_{\rm R23}$ than HSB galaxies. Both LSB and HSB galaxies show negative metallicity gradients across different mass ranges, with the gradient becoming steeper as stellar mass increases. Interestingly, the metallicity gradients of LSB galaxies are steeper than those of HSB galaxies. This may indicate that gas accretion has a more substantial impact on the outer regions of LSB galaxies. In the intermediate mass range (green lines), large-size galaxies demonstrate significantly higher 12+log(O/H)$_{\rm R23}$ than LSB galaxies, along with correspondingly gentler metallicity gradients.
    
    To check these relationships, we employed the PG16 and DOP16 methods, which use the [N$\,\textsc{ii}$] emission line to calculate the gas-phase metallicity (see the second and third rows of Fig. \ref{figA6}). The results align well with those derived from the R23 method, although notable differences arise in the low-mass range. In this mass range, the R23 method yields comparable 12+log(O/H) levels for LSB and HSB galaxies (blue line). In contrast, the PG16 and DOP16 methods reveal that while the central 12+log(O/H) of LSB galaxies is comparable to that of HSB galaxies, their peripheral metallicity values are significantly lower. This indicates that in the low-mass range, LSB galaxies have lower peripheral 12+log(O/H) and reduced [N$\,\textsc{ii}$] line strength relative to HSB galaxies, and they are predominantly classified as star-forming regions in the BPT diagram. Moreover, large-size galaxies show relatively higher 12+log(O/H) and a flatter metallicity gradient compared to their LSB counterparts.

    Overall, the gas-phase metallicity results obtained through various methodologies consistently indicate that LSB galaxies show lower gas-phase metallicity than both HSB and large-size galaxies, accompanied by marginally steeper negative metallicity gradient. These findings may imply that gas accretion plays a critical role in shaping the metallicity profiles of LSB galaxies, contributes to the steeper negative gradient observed in their gas-phase metallicity.
    
    \subsubsection{Radial Profile and Gradients of Stellar Population}
    \label{sec:Radial Profile and Gradients of Stellar Population}
    
\begin{figure*}[htbp]%
    \centering
    \includegraphics[width=0.8\textwidth,height=0.55\textheight]{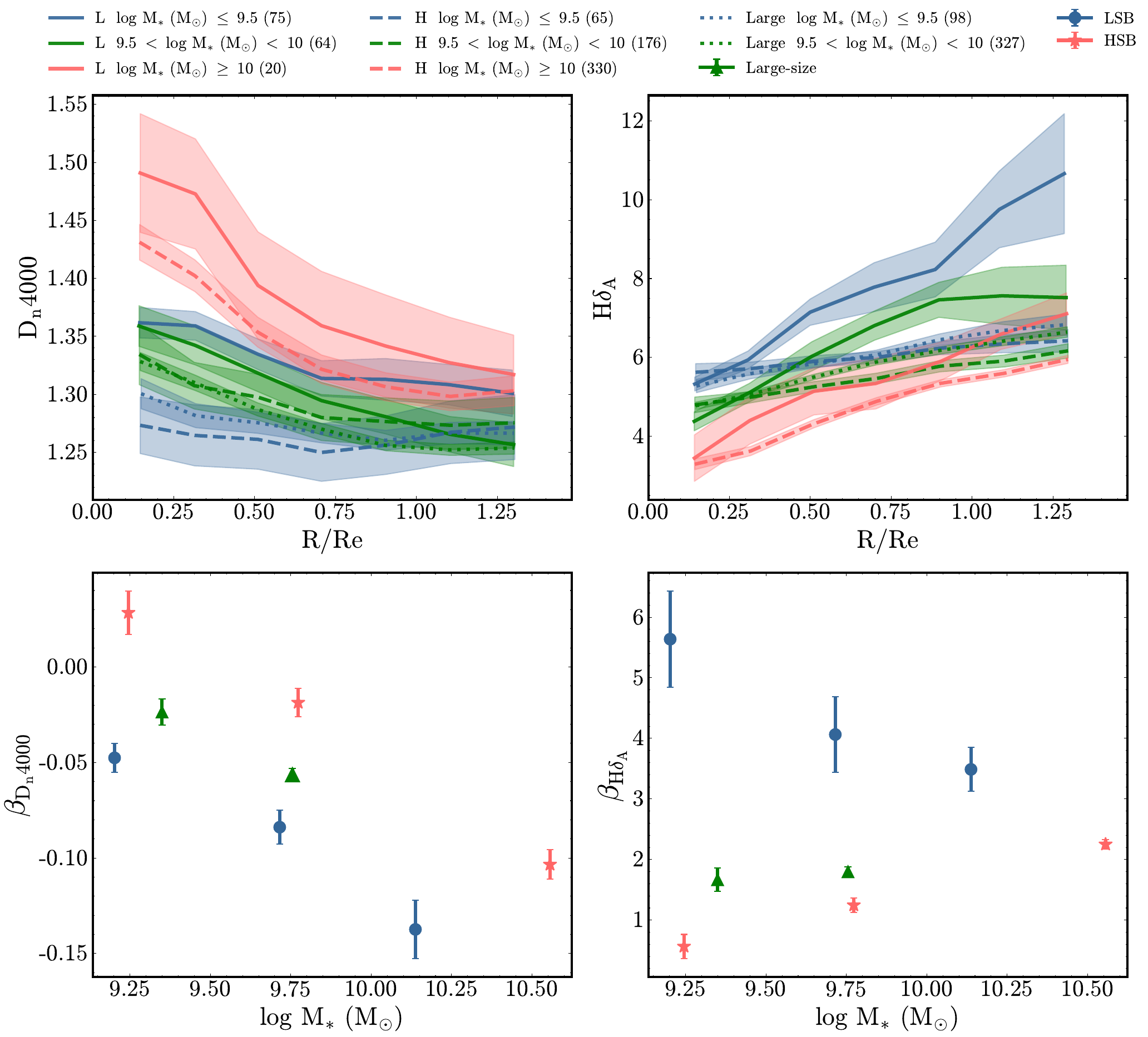}
    \caption{Radial profiles (top row) and median gradients (bottom row) of $D_{n}$4000 and H$\delta_A$ across different stellar mass ranges. The color/line conventions for different mass ranges is the same as in Fig. \ref{fig8}.}
    \label{fig11}
    \end{figure*}

    We extended our investigation into the stellar population characteristics of galaxies by analyzing their radial profiles and gradients of stellar age, specifically using two key spectral indices as diagnostic tracers: $D_{n}$4000 break and H$\delta_A$ absorption-line index. The $D_{n}$4000 index serves as a crucial indicator of the luminosity-weighted age of the stellar population, measuring the flux density ratio between two narrow spectral bands: 3850-3950\AA\; and 4000-4100\AA\; \citep{Bruzual_2003}. This index effectively captures the formation history of young stars in galaxies; a larger $D_{n}$4000 value correlates with an older average age of the stars in a galaxy. The H$\delta_A$ index, defined as the equivalent width of the H$\delta_A$ absorption feature within the 4083 - 4122\AA\; \citep{Worthey_1997}. Elevated H$\delta_A$ index values ($\geq$ 6\AA) trace enhanced recent star formation (within the past $\sim$1-2 Gyr), making it a robust proxy for younger stellar populations \citep{Kauffmann_G_2003}. For comparative purposes, Fig. \ref{fig11} illustrates the radial profiles and gradients of both $D_{n}$4000 and H$\delta_A$ indices for LSB, HSB and large-size galaxies in different mass ranges, the line/shape conventions are consistent with those adopted in Fig. \ref{fig8}. Fig. \ref{figA7} in Appendix \ref{Individual galaxies, individual bulge components and median profiles in different mass ranges} presents radial profiles of luminosity-weighted and mass-weighted stellar ages for LSB and HSB galaxies—these profiles independently validate the radial age trends inferred from $D_n4000$ and H$\delta_A$ indices, thus strengthening the robustness of our conclusions on stellar population age stratification in LSB and HSB galaxies.
    
    The first column of Fig. \ref{fig11} displays the radial profile and gradient of $D_{n}$4000, facilitating direct comparisons of galaxy age distributions. Our findings reveal that $D_{n}$4000 increases gradually with stellar mass, with LSB galaxies exhibiting higher values than HSB galaxies, particularly pronounced in low-mass systems. This suggests a general trend of older ages in LSB galaxies. In examining the radial profiles, we observe that LSB galaxies show a negative $D_{n}$4000 gradient, which becomes increasingly negative with rising stellar mass. Conversely, HSB galaxies transition from a positive gradient to a negative one. Overall, LSB galaxies display a steeper negative $D_{n}$4000 gradient, indicative of significant age stratification: older stellar populations dominate their central regions, while younger populations are concentrated in the outskirts. Furthermore, the $D_{n}$4000 of large-size galaxies is significantly lower than that of LSB galaxies, especially in the low-mass range, and has slightly elevated $D_{n}$4000 compared to HSB galaxies, albeit with marginally more negative gradients. These trends are further corroborated by the complementary analysis of luminosity- and mass-weighted ages presented in Fig. \ref{figA7}.

    The radial profile and gradient of H$\delta_A$ are presented in the second column of Fig. \ref{fig11}. Notably, LSB galaxies show significantly higher H$\delta_A$ values compared to HSB and large-size galaxies, particularly in low-mass galaxies, where this difference is most pronounced in their outskirts. Regarding the radial gradient, LSB galaxies experience a gradual decrease in their positive H$\delta_A$ gradient as stellar mass increases, whereas HSB and large-size galaxies show a steady increase in their positive gradients. This divergence results in a marked distinction in radial gradients between LSB galaxies and their HSB/large-size counterparts, especially within the low-mass range. The persistently steeper H$\delta_A$ gradients in LSB galaxies indicate a pronounced enhancement of this index in their outer regions, consistent with recent star formation activity dominated by an A-type stellar population.
    
    In general, LSB galaxies show higher $D_{n}$4000 and H$\delta_A$ compared to HSB and large-size galaxies. Their $D_{n}$4000 radial gradient is notably more negative, while the positive gradient of H$\delta_A$ is steeper. This contrast implies that their central regions may host older stellar populations, while ongoing gas accretion could be promoting recent A-type star formation in their outskirts. This interplay contributes to the observed steep radial gradients in both age indicators.
    
    \section{DISCUSSION}
    \label{sec:DISCUSSION}
    
    \subsection{Distinct Recent Star Formation Histories of LSB Galaxies}
    \label{sec:Distinct Recent Star Formation Histories of LSB Galaxies}

    To test whether the radial profiles of LSB galaxies are independent of $M_\ast$ and SFR, we analyze a control sample matched in these parameters. Fig. \ref{figA3} in Appendix \ref{The results of the control sample} shows that, even after matching, LSB galaxies retain radial trends similar to those of the original sample across all stellar mass ranges. Compared with HSB galaxies, LSB galaxies have suppressed values in $\Sigma_\ast$, $\Sigma_{\rm SFR}$, sSFR, and 12+log(O/H)$_{\rm R23}$. Importantly, their radial gradients also differ markedly from HSB counterparts. LSB galaxies show flat negative $\Sigma_{\rm SFR}$ gradients, steep positive sSFR gradients, and steep negative 12+log(O/H)$_{\rm R23}$ gradients.
    
    We further observe mass-dependent age stratification – elevated central D$_n$4000 in low-mass LSB galaxies and outer-disk youth signatures (reduced D$_n$4000) in intermediate and massive ranges. The outer H$\delta_A$ is generally enhanced, peaking in low-mass galaxies. Their radial gradients reveal accelerated negative D$_n$4000 declines and amplified positive H$\delta_A$ slopes, indicating older centers with starbursting peripheries. This pattern suggests that LSB evolution may be governed by dark matter-driven gas accretion sustaining extended star formation, independent of global M$_\ast$-SFR scaling.
    
    We also examine the star formation history (SFH) using the $D_n$4000-H$\delta_A$ diagnostic. A notable negative correlation exists: younger galaxies typically show deeper $H\delta$ absorption and weaker 4000\AA\; discontinuity. This interplay between the two indicators sheds light on recent star formation events, allowing us to determine whether these events have occurred continuously or in short bursts. 
    
\begin{figure*}[htbp]%
    \centering
    \includegraphics[width=0.98\textwidth,height=0.18\textheight]{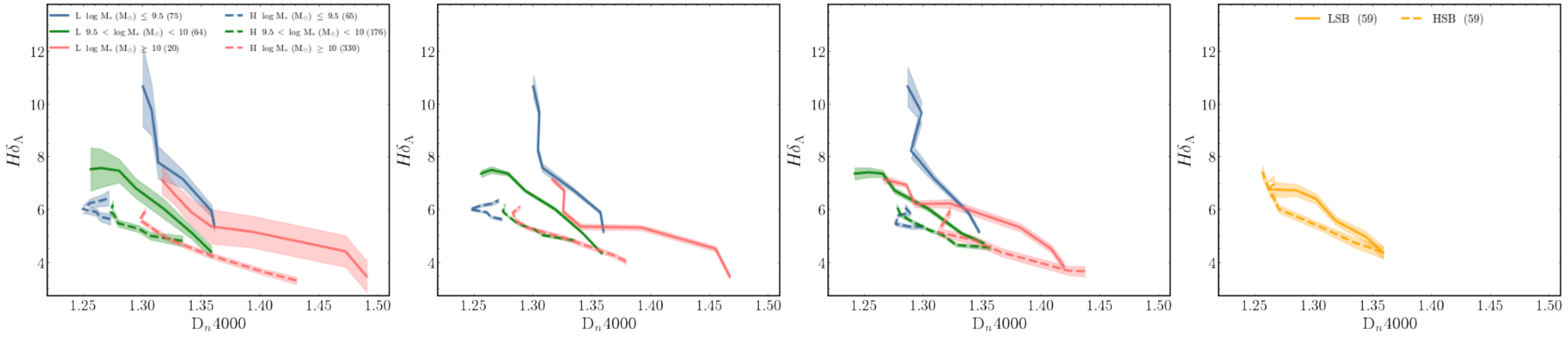}
    \caption{The relationship between $D_{n}$4000 and $H\delta_{A}$, where the color coding for the different stellar mass ranges follows the same scheme as in Fig. \ref{fig7}. The first panel shows the results for the original sample, while the second panel presents results for a mass-matched sample. The third panel illustrates results matched for both $M_\ast$ and SFR, and the fourth panel displays results for a mass-SFR-size-matched sample. Due to the limited number of galaxies satisfying the mass-SFR-size matching criteria, the fourth panel is not binned by stellar mass.}
    \label{fig12}
    \end{figure*}

    Fig. \ref{fig12} presents the $D_{n}$4000-H$\delta_A$ relation for four samples with progressively stricter matching criteria. The leftmost panel shows the original, unmatched sample, binned by stellar mass and colored following the scheme in Fig. \ref{fig7}. Across all stellar mass ranges, LSB galaxies exhibit systematically larger $D_{n}$4000 and H$\delta_A$ values than HSB galaxies. At fixed $D_{n}$4000, LSB galaxies show significantly higher H$\delta_A$, with the strongest offset occurring in the low-mass range, indicating experienced sustained star formation over the past $\sim$1–2 Gyr.
    
    The middle panels present samples matched in $M_\ast$ and in both $M_\ast$ and SFR, respectively, in order to control for the effects of galaxy mass and instantaneous star formation activity. In both cases, the offset persists: at fixed $D_{n}$4000, LSB galaxies consistently show higher H$\delta_A$ values than their HSB counterparts. This demonstrates that the observed SFH differences are not driven by variations in stellar mass or global SFR.
    
    The rightmost panel further matches galaxies in $R_e$ to account for potential size-related effects. Because of the stringent matching in $M_\ast$, SFR, and $R_e$, only 40 valid LSB–HSB pairs are retained, and no stellar mass subdivision is applied. Despite these constraints, LSB galaxies still show systematically higher H$\delta_A$ in their outskirts at fixed $D_{n}$4000. This persistence indicates that the divergent SFHs of LSB and HSB galaxies are intrinsic, rather than driven by differences in mass, SFR, or size. 
    
    To analyze whether recent star formation events in LSB galaxies occurred continuously or in bursts, we stacked the spectra of $D_n$4000 and H$\delta_A$ at the galactic center (0-1 $R_e$) and outer regions (1-2 $R_e$), along with the $u-g$ and $g-z$ colors, as shown in Appendix \ref{The stellar population and colors in the inner and outer regions of galaxies}. The correlation diagram is presented in Fig. \ref{figA4}, which reveals that LSB galaxies tend to exhibit enhanced star formation in their outskirts compared to their central regions. Most LSB galaxies (blue circles) display $D_n$4000 vs H$\delta_A$ and $u-g$ vs. $g-z$ color distributions similar to those of typical star-forming galaxies, with no significantly larger scatter. This indicates a continuous SFH rather than the bursty histories as predicted by simulations \citep{Boissier_2003}. Notably, some LSB galaxies show unusually high H$\delta_A$ values; however, we will defer a detailed investigation of their SFHs and spectral indices to future work.

    The sample was constructed by selecting galaxies with dominant disk components (B/T$<$0.66) as determined through multi-component photometric decomposition. While bulge components are included in the overall property study, we present the radial profiles (black solid lines) of the bulge components for LSB and HSB galaxies across various parameters in Appendix \ref{Individual galaxies, individual bulge components and median profiles in different mass ranges}, where the outer gray area corresponds to the disk-dominated region. These illustrations demonstrate that the MaNGA spectra encompasses both bulge and disk components in LSB and HSB galaxies, confirming the consistency between the characteristics of the outer disk component and the overall galaxy properties.

    %constrained to the same $M_\ast$
    
    \subsection{Evolution Patterns of LSB Galaxies}
    \label{sec:Evolution Patterns of LSB Galaxies}
    
    Global properties and radial profile analysis based on MaNGA spectral data reveal significant differences between LSB and HSB galaxies. Compared to large-size star-forming galaxies studied by \citet{Lin_2024}, LSB galaxies retain distinct properties that are not fully explained by their $M_\ast$ and size alone, suggesting that their evolutionary mechanisms are likely unique.
    
    The formation and evolution of LSB galaxies can be traced back to various factors such as dark matter halo properties, angular momentum distribution and environmental conditions. Observational and simulation studies suggest that LSB galaxies typically reside in isolated and low-density environments \citep{Galaz_2011, Du_2015, Honey_2018, Perez_Montano_2019}. These galaxies often inhabit dark matter halos with high spin parameters \citep{Boissier_2003, Rong_2017, Posti_2018} and continue to accrete primordial gas with high angular momentum through cold flows. This process leads to a gradual increase in mass, facilitating cooling and the development of extended disk structures. The high angular momentum \citep{Mo_1998, Perez_Montano_2019, Perez_Montano_2022} inhibits gas collapse toward the galactic center, resulting in low disk density, dispersed star formation, and low metallicity \citep{McGaugh_1994, Galaz_2006, Liang_2010}. The observed enhancement of H$\delta_A$ in LSB outskirts suggests that local gas accretion may trigger short star formation activity; however, the low metallicity limits gas cooling efficiency, preventing sustained high star formation. In contrast, HSB galaxies exist within dark halos that have low spin parameters and are situated in denser environments, such as galaxy clusters. Their low angular momentum leads to rapid gas collapse, forming dense disks where efficient star formation occurs. This process triggers metal enrichment and can lead to violent feedback, such as supernova explosions. Additionally, environmental interactions may further facilitate radial gas flow and bulge build-up in HSB galaxies. Consequently, LSB and HSB galaxies may evolve along divergent evolutionary paths.
    
    Research conducted by \citet{Ma_2024}, using TNG50 simulations of 836 disk galaxies, further elucidates the intrinsic evolution of compact and large-size galaxies. Their findings indicate that large-size galaxies grow in size through the continuous accretion of high angular momentum gas, exhibiting characteristics highly consistent with those of LSB galaxies, such as high gas content, low metallicity, and extended structures. In contrast, compact galaxies (analogous to HSB galaxies) develop dense structures due to rapid early accretion. While LSB galaxies display a flatter negative gradient in $\Sigma_{\rm SFR}$ and a steeper negative gradient in 12+log(O/H) compared to these large-size galaxies, with significantly lower outer $D_n$4000 and higher outer H$\delta_A$, their inner and outer regions occupy overlapping positions in both the $D_n$4000 vs. H$\delta_A$ and $u-g$ vs. $g-z$ color diagrams. This suggests that LSB galaxies may share an accretion-dominated evolutionary path with large-size galaxies, albeit evolving at a slower rate.
    
    \section{SUMMARY AND CONCLUSION}
    \label{sec:SUMMARY AND CONCLUSION}
    
    In this paper, we analyzed the properties of 1,223 face-on late-type galaxies without AGN from SDSS-MaNGA DR17. We performed \textsc{SerExp} and \textsc{SerSer} disk/bulge component decompositions on $g$-band images from the DESI Legacy Survey, yielding reliable results for 1,118 galaxies. For these galaxies, we calculated the corrected central surface brightness $\mu_{\rm 0,d,cor}(g)$ of their disk components in $g$ band. Using the selection criterion $\mu_{\rm 0,d,cor}(g) = 22 \pm 0.3$ mag arcsec$^{-2}$, we classified the galaxies into 159 LSB galaxies ($\mu_{\rm 0,d,cor}(g) \geq$22.3 mag arcsec$^{-2}$), 388 LSB candidates (21.7$< \mu_{\rm 0,d,cor}(g) <$22.3 mag arcsec$^{-2}$), and 571 HSB galaxies ($\mu_{\rm 0,d,cor}(g) \leq$21.7 mag arcsec$^{-2}$). 
    
    We further investigated the global properties of LSB galaxies, as well as their radial profiles and gradients, using IFU data from the MaNGA survey. Below are the key findings from our comparison between LSB and HSB galaxies across the parameter space defined by total stellar mass, SFR, and effective radius:
    
    \begin{itemize}
    \item LSB galaxies predominantly have stellar masses below $3 \times$ $10^{10}$ M$_\odot$, with only three galaxies exceeding this threshold. Compared to their HSB counterparts, LSB galaxies have systematically lower star formation rates and gas-phase metallicities, and possess larger effective radii.

    \item Across stellar mass ranges, we quantify systematic offsets ($\Delta$) from well-established galaxy-wide mass-scaling relations \citep{Belfiore_2018,Tremonti_2004,Shen_2003} in typical star-forming galaxies: $\Delta \log(\mathrm{SFR})$, $\Delta 12 + \log(\mathrm{O}/\mathrm{H})$ and $\Delta \log(R_e)$. More massive galaxies show reduced $\Delta \log(\mathrm{SFR})$ and $\Delta 12 + \log(\mathrm{O}/\mathrm{H})$, but enhanced $\Delta \log(R_e)$, indicating suppressed star formation, lower gas-phase metallicity, and larger effective radii. Notably, LSB galaxies exhibit distinct systematic offsets: depressed $\Delta \log(\mathrm{SFR})$ and $\Delta 12 + \log(\mathrm{O}/\mathrm{H})$ alongside elevated $\Delta \log(R_e)$ across all mass ranges. These offsets show weak to moderate but statistically significant correlations with disk surface brightness ($p < 0.001$), with the strongest correlations found for radius offsets ($r = 0.5$-$0.7$, depending on the mass range). Metallicity offsets show weaker correlations ($r = 0.1$-$0.2$), significant only within the stellar mass range $\log(M_\ast/\rm M_\odot) =$ 9.5-10.
    
    \item Likewise, more massive galaxies show higher $D_n$4000 (a well-established proxy of luminosity-weighted stellar age, with higher values directly indicating an older, less actively star-forming stellar population), higher dust attenuation, and shorter gas depletion times. The galaxy-wide $D_n$4000 index and H$\,\textsc{i}$ gas depletion time exhibit weak but statistically significant correlations with disk surface brightness, with both correlation increasing moderately with stellar mass. Despite having similarly high H$\,\textsc{i}$ content and low nebular dust attenuation ($A_V = 0.2$--0.4), as expected for low-mass galaxies, LSB galaxies show higher $D_n$4000 values and longer gas depletion times compared to HSB galaxies of similar stellar mass.
    
    \item Resolved MaNGA data show that their cumulative stellar mass growth of LSB galaxies rises more gradually from the center outward compared to that of HSB and large-size galaxies, indicating that LSB galaxies build up the same total stellar mass over a larger spatial extent. Moreover, both their mass surface density and cumulative mass profiles differ significantly from those of large-size galaxies, reinforcing that LSB galaxies are not simply large, low-mass systems.
    
    \item The radial profiles of SFR and sSFR in LSB galaxies reveal centrally suppressed star formation, with steep declines within the inner $\sim$5 kpc and enhanced activity in the outskirts. Specifically, their central SFRs are a factor of $\sim$2-3 \textit{lower} than those of HSB galaxies. This results in a positive sSFR gradient, distinguishing LSB galaxies from both HSB and large-size galaxies and reflecting a distinct spatial distribution of star formation.
    
    \item LSB galaxies consistently show lower gas-phase metallicity values across all radii and steeper negative radial metallicity gradients compared to both HSB and large-size galaxies. This suggests ongoing gas accretion plays a role in shaping their chemical enrichment, a trend is robust across different metallicity calibration methods.
    
    \item Compared with HSB and star-forming large-size galaxies, LSB galaxies show elevated $D_n$4000 values and steeper negative gradients, indicating older stellar populations in their centers and younger populations in the outskirts. Globally, their H$\delta_A$ index is notably high ($H\delta_A > 6\,\text{\AA}$) with a steeper positive gradient, particularly in the low-mass galaxies. At fixed $D_n$4000 value, LSB galaxies maintain higher H$\delta_A$ indices, consistent with ongoing or recent star formation in their outer regions. Comparing the $D_n$4000 vs H$\delta_A$ and $u-g$ vs $g-z$ color diagrams in the inner (0-1 $R_e$) and outer (1-2 $R_e$) regions, LSB galaxies align with star-forming galaxies, suggesting a continuous, rather than a bursty, star formation history.
    
    \end{itemize}

    Overall, this study supports a theoretical picture in which LSB and HSB galaxies evolve along divergent paths, driven by differences in their dark matter halos, angular momentum, and environmental conditions. LSB galaxies typically reside in isolated, low-density environments with high spin parameters and continuous gas accretion; they are characterized by larger sizes, lower central star formation rates, and lower metallicities, coupled with enhanced star formation activity in their outer regions. Like large-size galaxies, LSB galaxies show a continuous (rather than a bursty) star formation history, implying that both types may follow a shared accretion-driven evolutionary path. However, the evolution of LSB galaxies appears to proceed at a slower pace. These differences emphasize the distinct evolutionary processes that govern the formation and evolution of LSB galaxies.
    
    In future work, we will investigate the multiscale environments and dark matter halo masses of LSB and HSB galaxies in detail to further test the physical scenarios proposed for their formation and evolution.
    
    \newpage
    
    \section*{Acknowledgement}

    We thank the anonymous referee for comments that helped to improve this paper. The authors wish to thank P{\'e}rez-Monta{\~n}o and Min Du for helpful discussions and P{\'e}rez-Monta{\~n}o for suggesting the continuous evolution of LSB galaxies. This work is supported by the National Natural Science Foundation of China (NSFC) under Nos. 12373007, 12422302. This work is supported by National Key R$\&$D Program of China No.2022YFF0503402, and the National Natural Science Foundation of China (NSFC, Nos. 12233005). J.L. acknowledges the supports from the National Natural Science Foundation of China (No. 12403016). LL acknowledges the supports from the Youth Innovation Promotion Association CAS (id. 2022260) and the China Manned Space Program with grant No.CMS-CSST-2021-B04. J.W. acknowledges NSFC grants 12033004, 12333002, 12221003 and the China Manned Space Program with grant no. CMS-CSST-2025-A10. SS thanks research grants from the China Manned Space Project with NO. CMS-CSST-2025-A07, National Natural Science Foundation of China (No. 12141302) and Shanghai Academic/Technology Research Leader (22XD1404200). YR acknowledges supports from the NSFC grant 12273037, the CAS Pioneer Hundred Talents Program (Category B), and the USTC Research Funds of the Double First-Class Initiative. This work is supported by the China Manned Space Program with grant no. CMS-CSST-2025-A06 and CMS-CSST-2025-A08. 
    
    This work makes use of data from SDSS-IV. Funding for SDSS-IV has been provided by the Alfred PSloan Foundation and Participating Institutions. Additional funding towards SDSS-IV has been provided by the US Department of Energy Office of Science. SDSS-IV acknowledges support and resources from the Centre for High-Performance Computing at the University of Utah. The SDSS web site is www.sdss.org.
    
    SDSS-IV is managed by the Astrophysical Research Consortium for the Participating Institutions of the SDSS Collaboration including the Brazilian Participation Group, the Carnegie Institution for Science, Carnegie Mellon University, the Chilean Participation Group, the French Participation Group, Harvard–Smithsonian Center for Astrophysics, Instituto de Astrofsica de Canarias, The Johns Hopkins University, Kavli Institute for the Physics and Mathematics of the Universe (IPMU)/University of Tokyo, Lawrence Berkeley National Laboratory, Leibniz Institut fur Astrophysik Potsdam (AIP), Max-Planck-Institut fur Astronomie (MPIA Heidelberg), Max-Planck-Institut fur Astrophysik (MPA Garching), Max-Planck-Institut fur Extraterrestrische Physik (MPE), National Astronomical Observatory of China, New Mexico State University, New York University, University of Notre Dame, Observatario Nacional/MCTI, The Ohio State University, Pennsylvania State University, Shanghai Astronomical Observatory, United Kingdom Participation Group, Universidad Nacional Autonoma de Mexico, University of Arizona, University of Colorado Boulder, University of Oxford, University of Portsmouth, University of Utah, University of Virginia, University of Washington, University of Wisconsin, Vanderbilt University and Yale University.
    
    \bibliography{ref}{}
    \bibliographystyle{aasjournal}
    
    \appendix
    \renewcommand\thefigure{\Alph{section}\arabic{figure}}
    \setcounter{figure}{0} 
    
    \section{Images of all selected LSB galaxies in the sample}
    \label{Images of all selected LSB galaxies in the sample}
    
\begin{figure}[htbp] 
    \centering
    \includegraphics[width=0.95\textwidth]{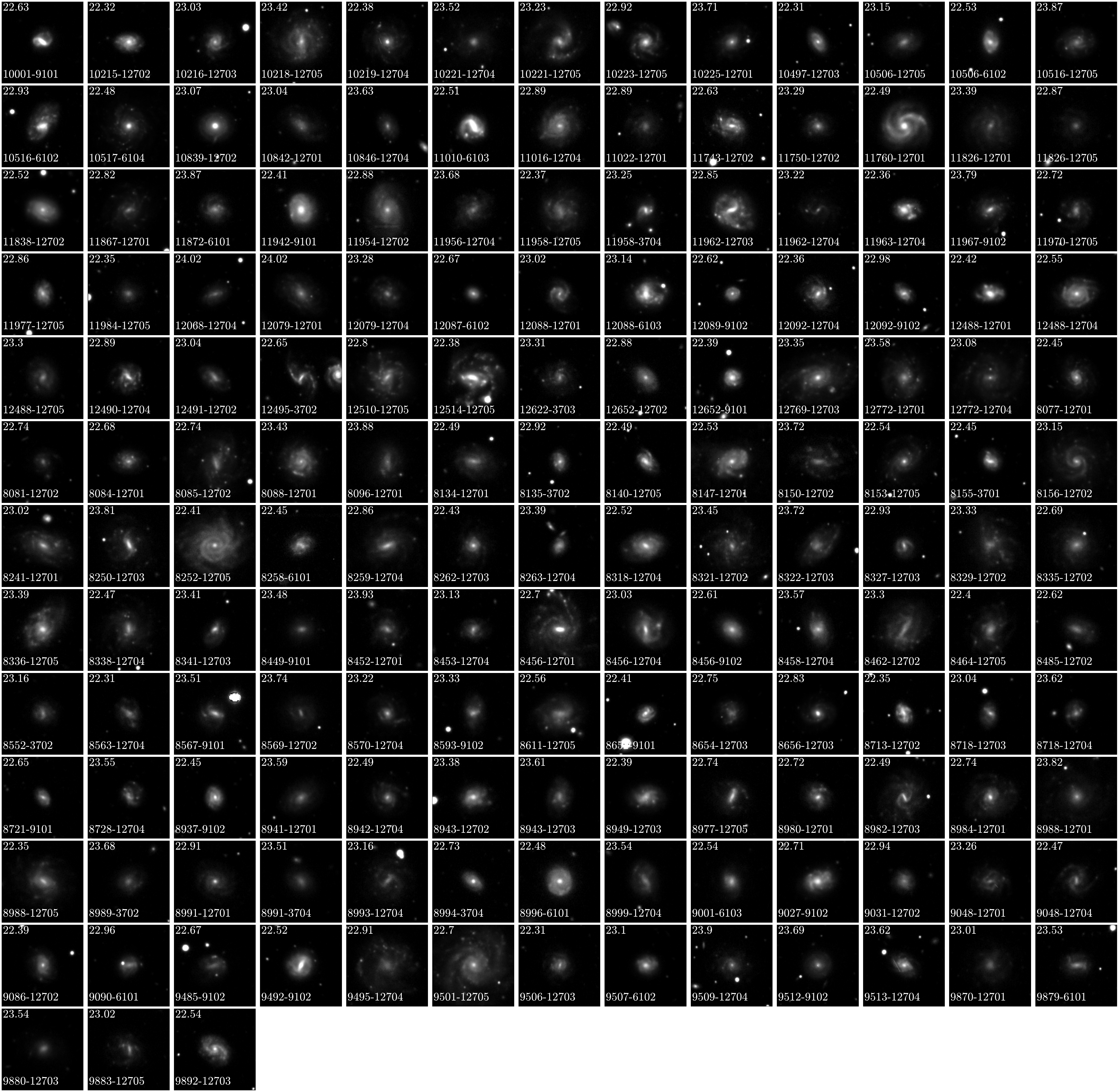} 
    \caption{The $g$-band photometric images of all 159 LSB galaxies in our sample. The corrected central surface brightness of the $g$-band disk component (in mag arcsec$^{-2}$) is labeled in the upper left corner of each panel.}
    \label{figA1}
    \end{figure}

    We present in Fig. \ref{figA1} the $g$-band photometric images of the LSB galaxies selected from our two-component decomposition. The upper left corner of each panel shows the corrected central surface brightness of the $g$-band disk component (in units of mag arcsec$^{-2}$) for the corresponding galaxy.
    
    \section{The results of the control sample}
    \label{The results of the control sample}
\begin{figure}[htbp] 
    \centering
    \includegraphics[width=0.80\textwidth]{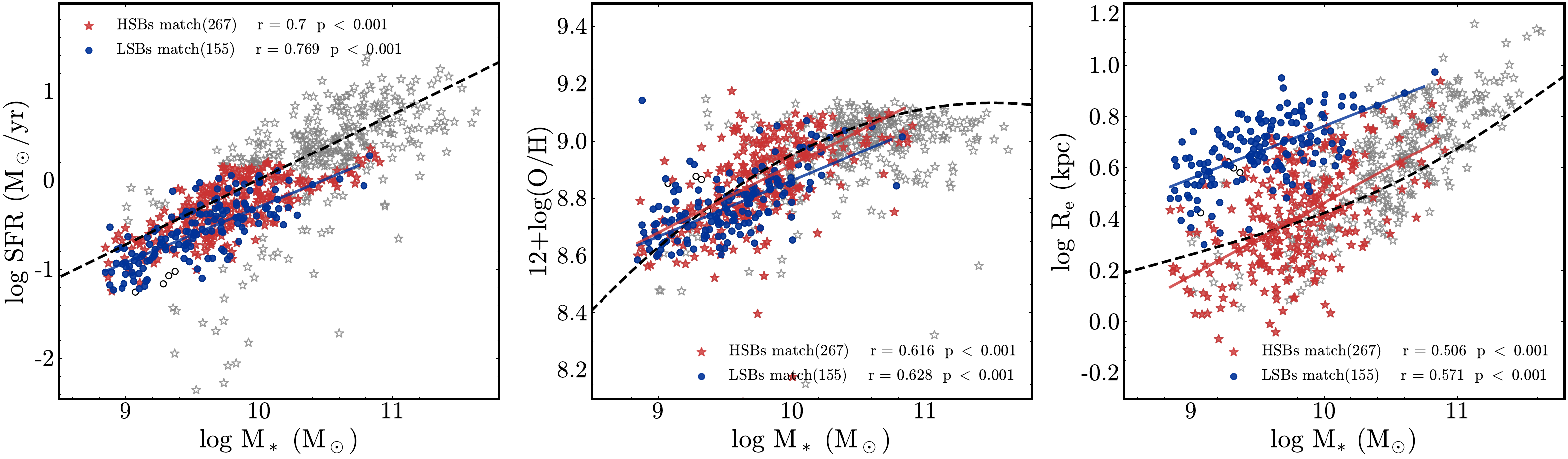} 
    \caption{The relationship between stellar mass, SFR, 12+log(O/H) and size for LSB (blue circles) and HSB galaxies (red stars) after matching $M_\ast$ and SFR. The black circles and gray stars represent unmatched LSB and HSB galaxies, respectively. The lines follow the same style as Fig. \ref{fig4}.}
    \label{figA2}
    \end{figure}

\begin{figure}[htbp] 
    \centering
    \includegraphics[width=0.80\textwidth]{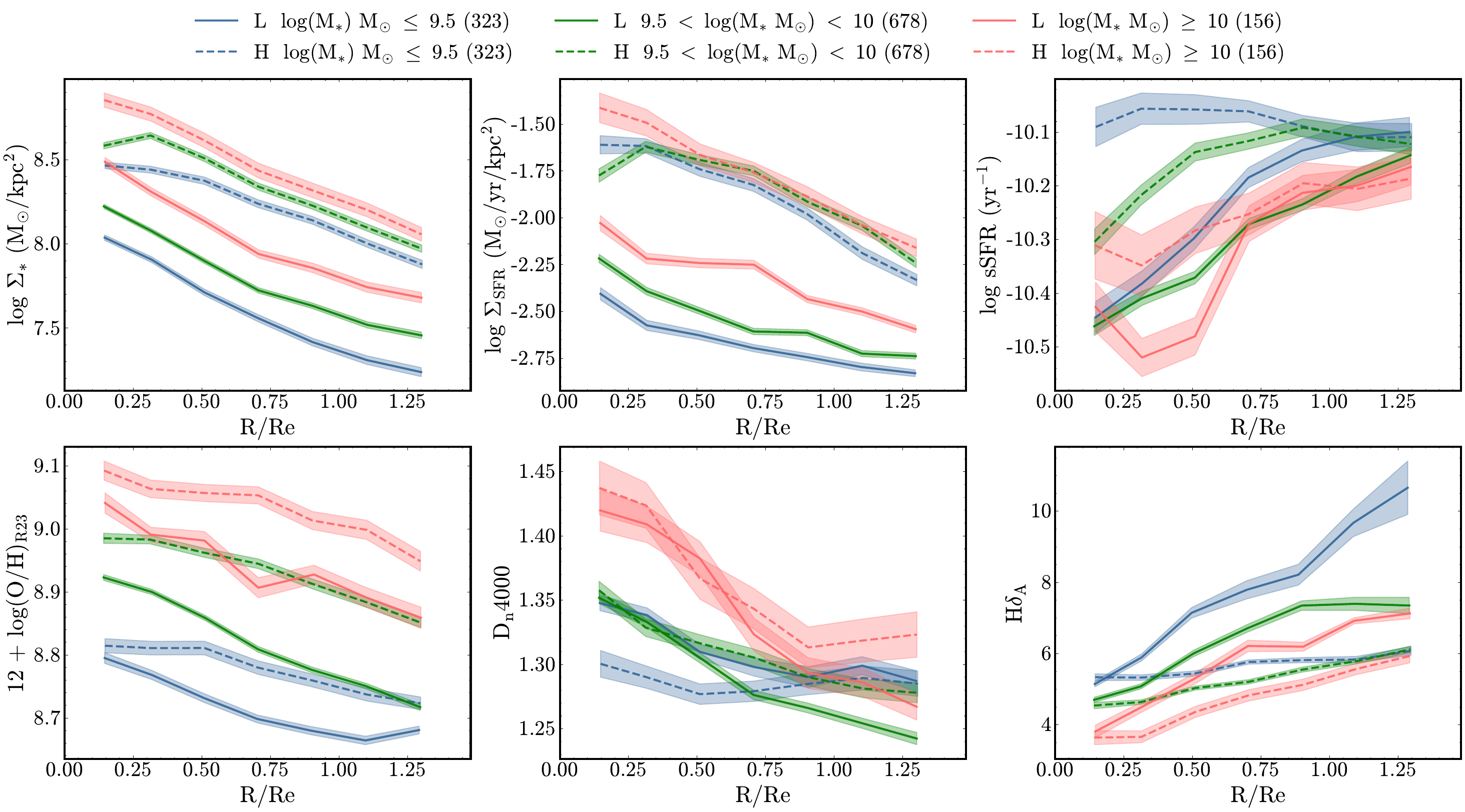} 
    \caption{Radial profiles (normalized to $R/R_e$) of $\Sigma_\ast$, $\Sigma_{\rm SFR}$, sSFR, 12+log(O/H)$_{\rm R23}$, $D_n$4000, and H$\delta_A$ for LSB (solid lines) and HSB galaxies (dashed lines) after matching $M_\ast$ and SFR. The lines follow the same style as Fig. \ref{fig7}.}
    \label{figA3}
    \end{figure}

    In Section \ref{sec:Control Sample of LSB and HSB Galaxies}, we matched LSB galaxies with their HSB counterparts by $M_\ast$ and SFR. This section presents the global properties of the matched sample (Fig. \ref{figA2}) and analyzes the radial profiles of key parameters ($\Sigma_{\ast}$, $\Sigma_{\rm SFR}$, sSFR, 12+log(O/H)$_{\rm R23}$, $D_n$4000 and H$\delta_A$) for LSB and HSB galaxies after quantitative weighting (Fig. \ref{figA3}).
    
    As illustrated in Fig. \ref{figA2}, even when matched by $M_\ast$ and SFR, LSB galaxies show lower 12+log(O/H) and significantly larger sizes compared to HSB galaxies. A closer examination of Fig. \ref{figA3} reveals that within different mass ranges, LSB galaxies consistently show lower $\Sigma_{\ast}$, $\Sigma_{\rm SFR}$, sSFR and 12+log(O/H)$_{\rm R23}$ than matched HSB galaxies. Notably, in the lower mass range, LSB galaxies have a relatively high $D_n$4000, while LSBs in the intermediate and massive ranges show similar central $D_n$4000 but lower $D_n$4000 values in their outskirts. Additionally, H$\delta_A$ is consistently higher in the outskirts of LSB galaxies across all mass ranges. Therefore, even after matching the global properties of $M_\ast$ and SFR, these findings highlight distinct spatial distribution characteristics between LSB and HSB galaxies, confirming that these spatial differences are independent of global stellar mass and star forming rate. Specifically, LSB galaxies show a gentle negative gradient in $\Sigma_{\rm SFR}$, a steeper positive gradient in sSFR, relatively steep negative gradients in $D_n$4000, and positive gradients in H$\delta_A$. This contrast underscores the significance of these spatial differences, providing valuable insights for future research.

    \section{The stellar population and colors in the inner and outer regions of galaxies}
    \label{The stellar population and colors in the inner and outer regions of galaxies}
    
    To investigate the star formation history of LSB galaxies, we plotted the H$\delta_A$ vs. $D_n$4000 relation, and $u-g$ vs $g-z$ colors diagram, for two radial regions: the inner region (0-1 $R_e$) and outer region (1-2 $R_e$). The figures reveal that large-size galaxies (gray contours), HSB galaxies (red stars), LSB candidates (black open triangles), and LSB galaxies (blue circles) show similar distributions in both the H$\delta_A$ - $D_n$4000 relation and color diagram. This similarity suggests that LSB galaxies, like large-size and HSB galaxies, have a continuous star formation history, though LSB galaxies show slower evolutionary rates.
    
    \begin{figure}[htbp] 
    \centering
    \includegraphics[width=0.55\textwidth]{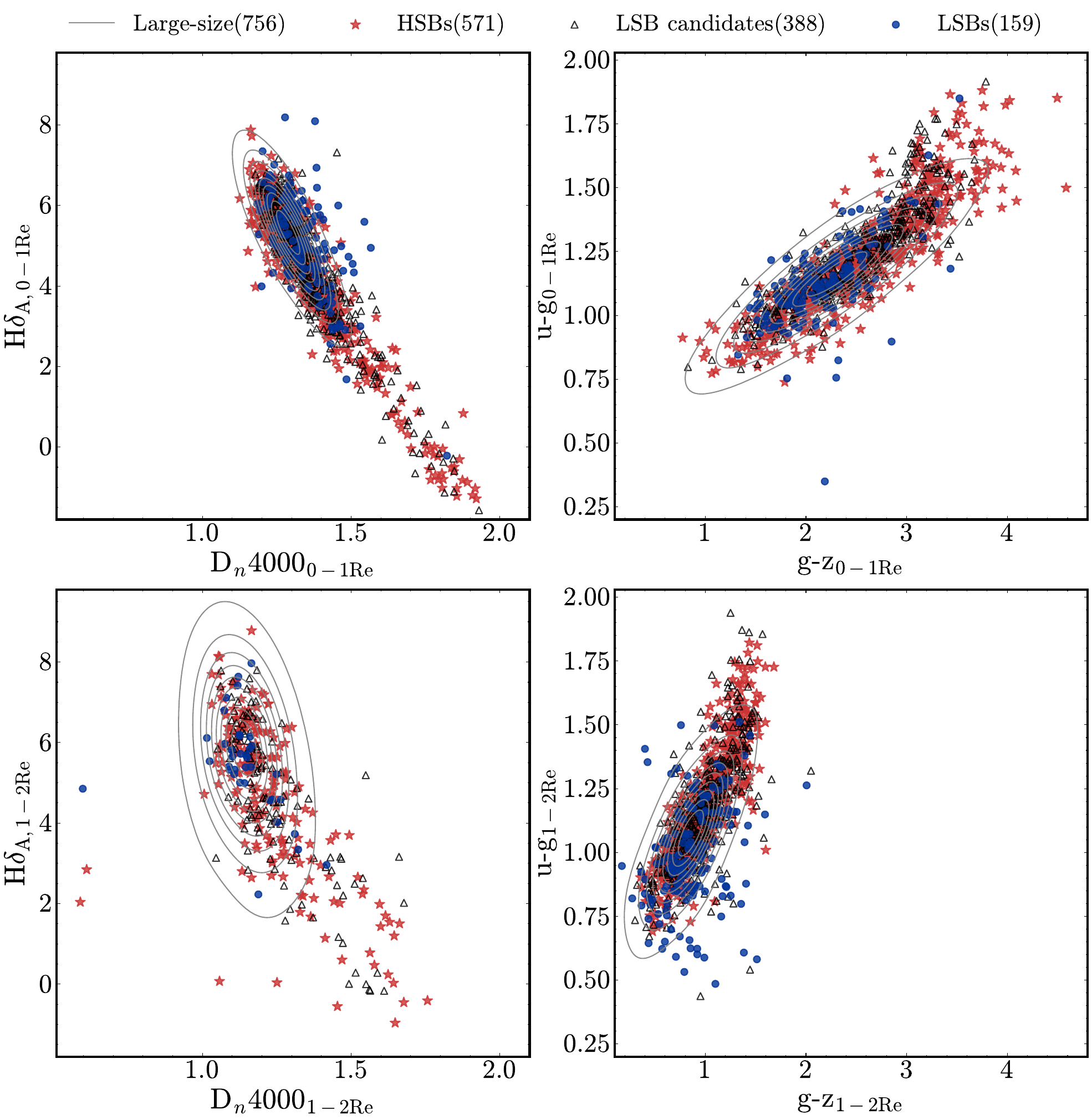} 
    \caption{Stellar population and color distribution across radial regions. The first column shows the relationship between H$\delta_A$ and $D_n$4000 from MaNGA data, while the second column displays the $u-g$ vs. $g-z$ color diagram obtained from SDSS images. The top row corresponds to the inner region (0-1 $R_e$), and the bottom row to the outer region (1-2 $R_e$). In these figures, blue circles represent LSB galaxies, hollow triangles denote LSB candidates, red stars indicate HSB galaxies, and gray contours outline the distribution of large-size galaxies.}
    \label{figA4}
    \end{figure}
    
    \section{Individual galaxies, individual bulge component profiles and median profiles in different mass ranges}
    \label{Individual galaxies, individual bulge components and median profiles in different mass ranges}
    
\begin{figure}[htbp] 
    \centering
    \includegraphics[width=0.7\textwidth]{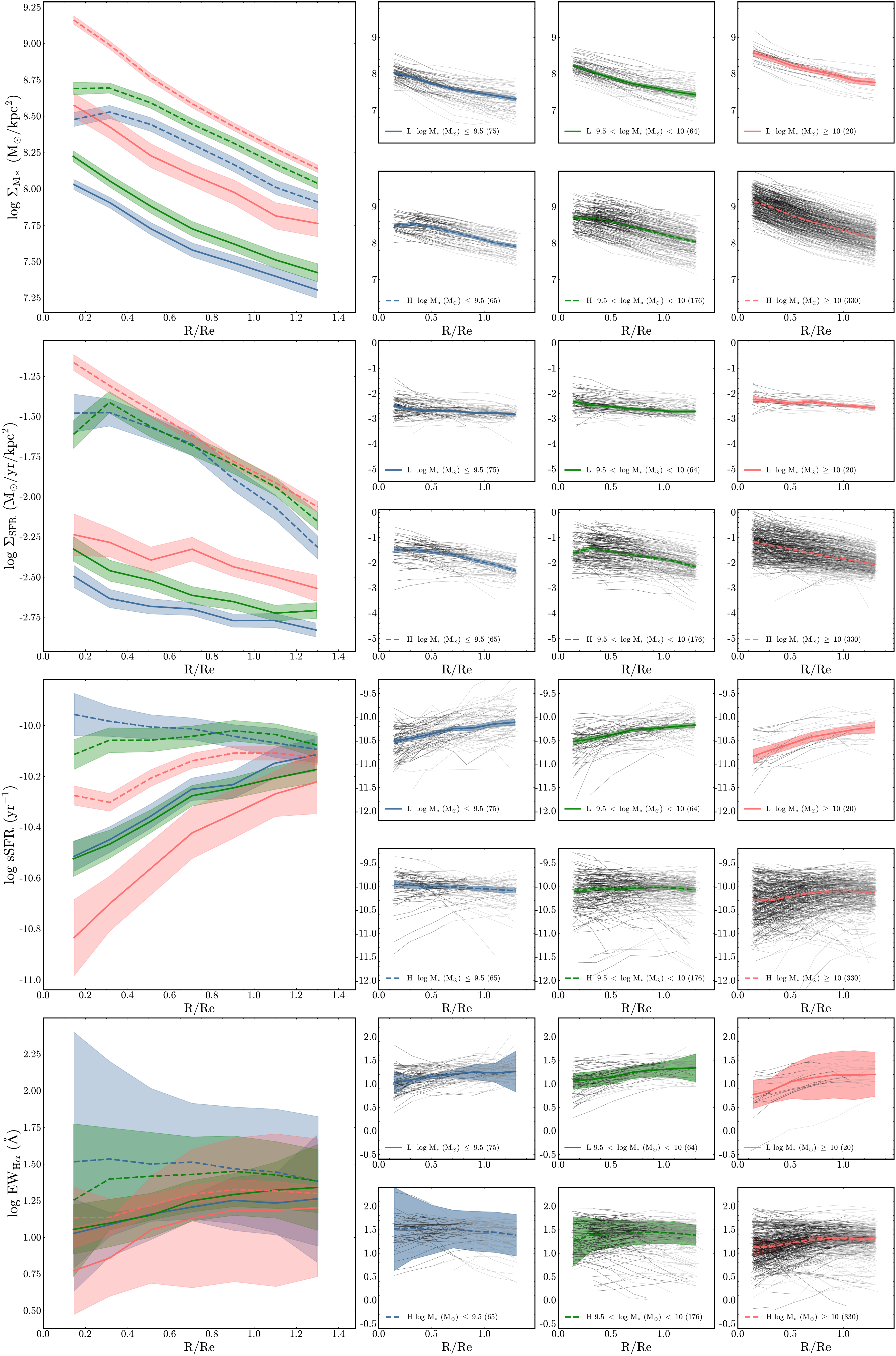} 
    \caption{Radial distribution of $\Sigma_{\ast}$, $\Sigma_{\rm SFR}$, sSFR, and $EW_{H\alpha}$ for various stellar mass ranges. Gray lines represent individual galaxy profiles, with the radius sampled in steps of 0.2 $R_e$ along the semi-major axis. Black lines indicate the location of the bulge for the same galaxy. Median profiles are color-coded by stellar mass range, blue (log($M_\ast/M_{\odot}$) $\leq$ 9.5), green (9.5 $<$ log($M_\ast/M_{\odot}$) $<$ 10), and red (log($M_\ast/M_{\odot}$) $\geq$ 10). Solid lines correspond to LSB galaxies, dashed lines to HSB galaxies. Shaded areas reflect the error of the median profile.}
    \label{figA5}
    \end{figure}

\begin{figure}[htbp] 
    \centering
    \includegraphics[width=0.79\textwidth]{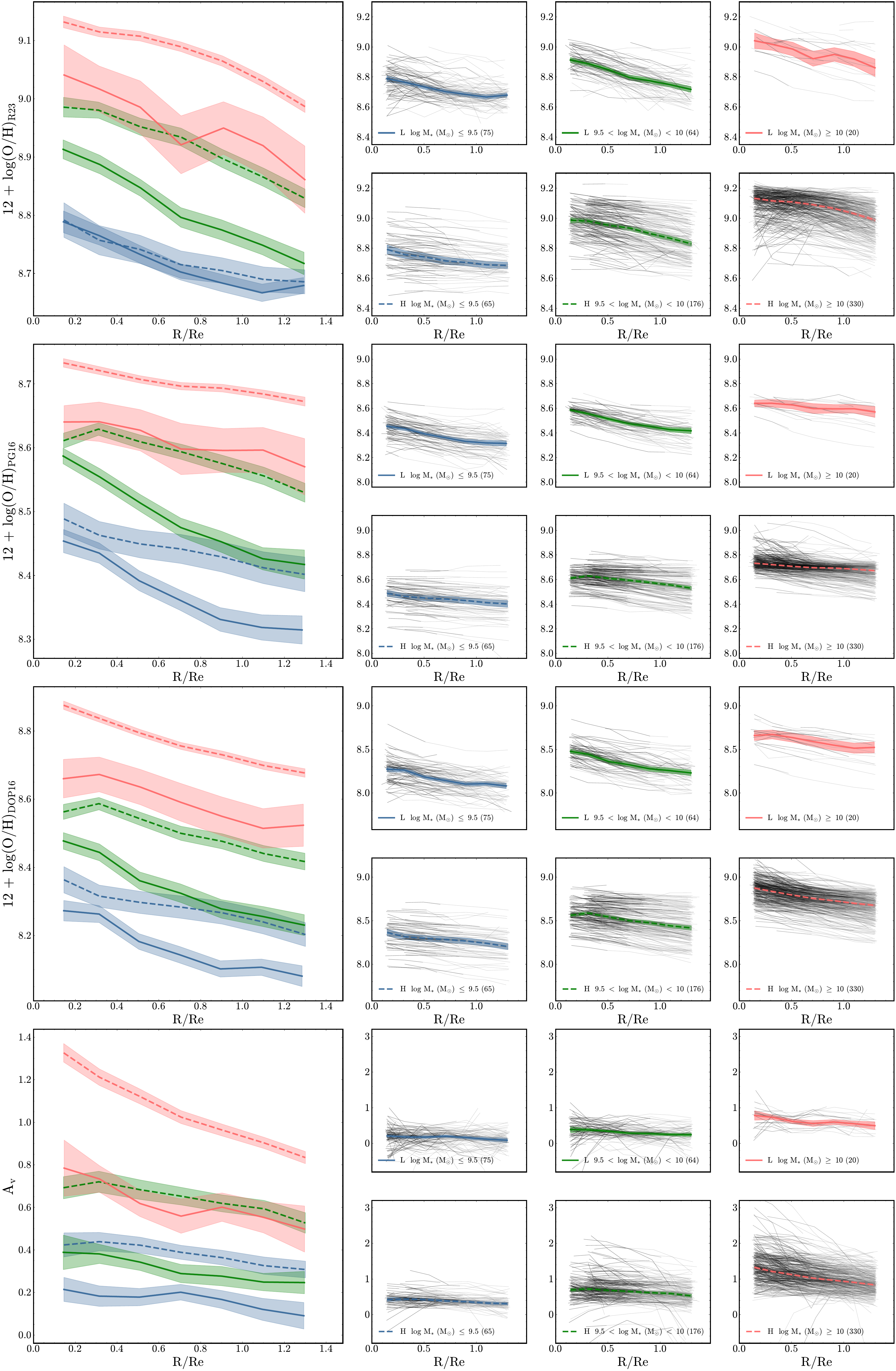} 
    \caption{Same as Fig. \ref{figA5}, but showing radial distribution of 12+log(O/H)$_{\rm R23}$, 12+log(O/H)$_{\rm PG16}$, 12+log(O/H)$_{\rm DOP16}$, and $A_{\rm V}$.}
    \label{figA6}
    \end{figure}

\begin{figure}[htbp]
    \centering
    \includegraphics[width=0.79\textwidth]{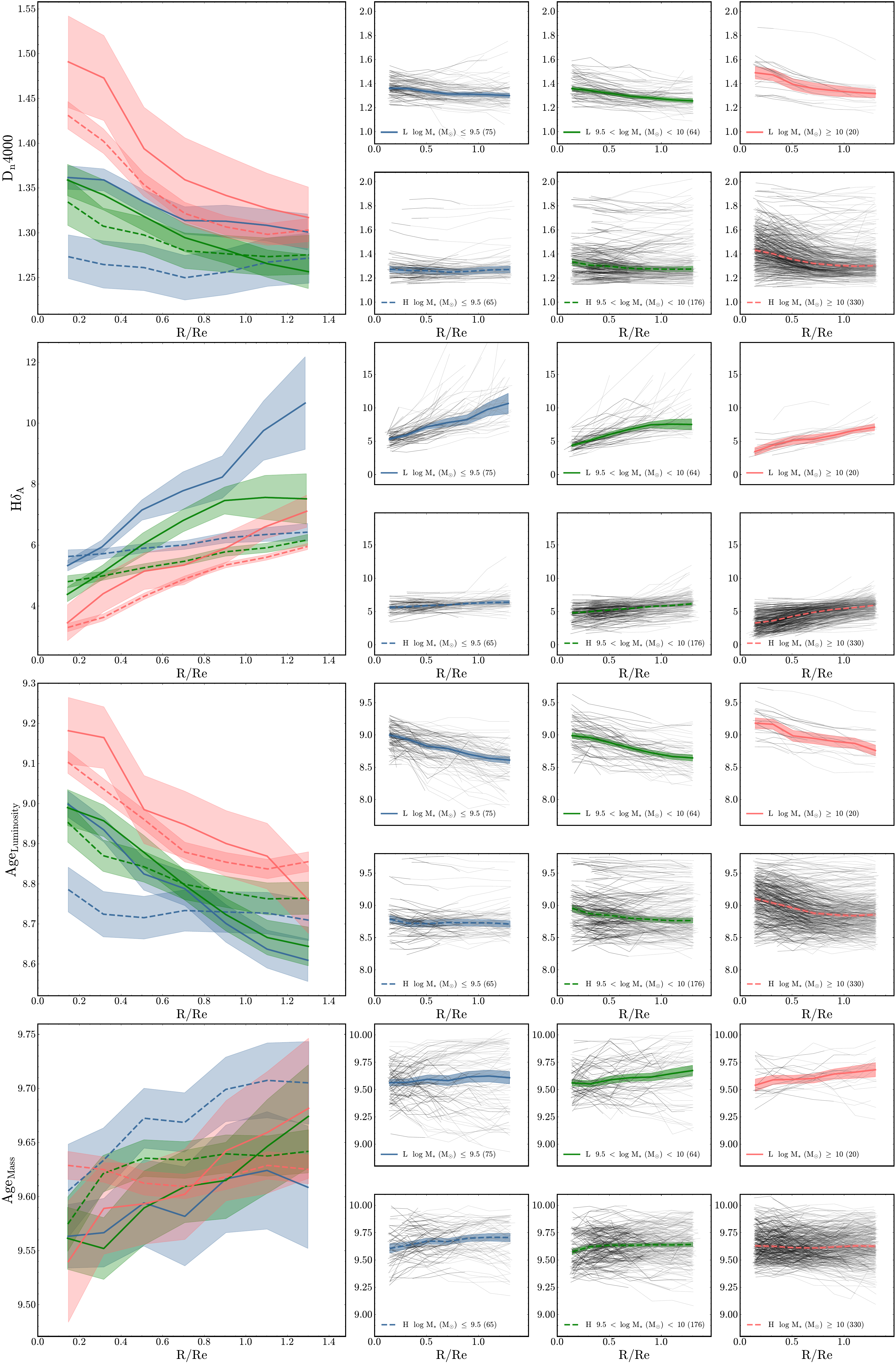}
    \caption{Same as Fig. \ref{figA5}, but showing radial distribution of $D_n$4000, H$\delta_A$, luminosity-weighted age Age$_{\rm luminosity}$, and mass-weighted age Age$_{\rm mass}$.}
    \label{figA7}
    \end{figure}

    Fig. \ref{figA5}-\ref{figA7} show radial profiles for LSB and HSB galaxies including individual profiles (gray lines), bulge component profiles (highlighted in black), and median profiles (color-coded by stellar mass range). The parameters analyzed here include: $\Sigma_{\ast}$, $\Sigma_{\rm SFR}$, sSFR, $EW_{H\alpha}$, 12+log(O/H)$_{\rm R23}$, 12 +log(O/H)$_{\rm PG16}$, 12 +log(O/H)$_{\rm DOP16}$, $A_{V}$, $D_n$4000, H$\delta_A$, Age$_{\rm Luminosity}$ and Age$_{\rm Mass}$. These radial profiles are derived from two-dimensional maps using position angle (PA) and inclination value from the NSA photometric catalog.
    
    Previous studies have focused on the global properties of LSB galaxies; here we use MaNGA spectral data to investigate their radial gradients. Since LSB galaxies are defined by the central surface brightness of their disk components, our analysis prioritizes disk properties. Therefore, it is crucial to consider the MaNGA spectra coverage, as the aperture effect may lead to incomplete coverage of the entire galaxy or over-concentration of data in the bulge region.
    
    The radial profiles (black lines) of the bulge components of Fig. \ref{figA5}-\ref{figA7} show that, for most LSB and HSB galaxies across different mass ranges, the MaNGA spectra cover not only the bulge component but also portions of the disk. Moreover, the characteristics of the outer disk components align with the conclusions from the global radial profile analysis in Section \ref{sec:Radial Profiles and Gradients}, validating the scientific value of studying LSB galaxies' radial profiles and gradients.%making the study of the radial profile and gradient of LSB galaxies meaningful.%

\end{CJK}
\end{document}